\documentclass[showpacs,twocolumn,superscriptaddress,pra,longbibliography,%
floatfix]{revtex4-1}

\usepackage{epsf}
\usepackage{amsmath}
\usepackage{amsfonts}
\usepackage{amssymb}
\usepackage{bm}
\usepackage{bbm}
\usepackage{nicefrac}
\usepackage{cancel}
\usepackage{xcolor}
\usepackage{maybemath}
\usepackage{pifont}
\usepackage{afterpage}
\usepackage{subfloat}
\usepackage{siunitx}

\usepackage{dcolumn}

\newcolumntype{.}{D{x}{}{-1}}

\usepackage{graphicx}
\usepackage{epsfig}
\usepackage{latexsym}
\usepackage{dcolumn}
\usepackage{pifont}

\definecolor{kellygreen}{rgb}{0.3, 0.73, 0.09}

\definecolor{garrosgreen}{rgb}{0.1, 0.4, 0.1}
\definecolor{dartmouthgreen}{rgb}{0.05, 0.5, 0.06}

\definecolor{duelferred}{rgb}{0.7, 0.2, 0.1}
\definecolor{cambridgeblue}{rgb}{0.1, 0.3, 1.0}
\definecolor{oxfordblue}{rgb}{0.05, 0.2, 0.7}

\def\ii{{\mathrm{i}}}

\def\vdw{van der Waals}

\def\HFS{{\mathrm{HFS}}}

\def\LS{{\mathrm{LS}}}
\def\FS{{\mathrm{FS}}}
\def\vdW{{\mathrm{vdW}}}

\def\calH{{\mathcal H}}
\def\calL{{\mathcal L}}

\def\calO{{\mathcal O}}
\def\calV{{\mathcal V}}

\makeatletter

\newcommand{\Rmnum}[1]{\expandafter\@slowromancap\romannumeral #1@}
\makeatother

\newcolumntype{.}{D{x}{}{-1}}

\begin{document}

\newcommand{\addrROLLA}{Department of Physics,
Missouri University of Science and Technology,
Rolla, Missouri 65409-0640, USA}

\newcommand{\addrHDphiltheo}{Institut f\"ur Theoretische Physik,
Universit\"{a}t Heidelberg,
Philosophenweg 16, 69120 Heidelberg, Germany}

\newcommand{\addrLEBEDEV}{P. N. Lebedev Physics
Institute, Leninsky prosp.~53, Moscow, 119991 Russia}

\newcommand{\addrMUC}{Max--Planck--Institut f\"ur 
Quantenoptik, Hans--Kopfermann-Stra\ss{}e~1, 
85748 Garching, Germany}

\newcommand{\addrRQC}{Russian Quantum Center, 
Business-center ``Ural'', 100A Novaya street, 
Skolkovo, Moscow, 143025 Russia}

\newcommand{\addrBUDKER}{Budker Institute of Nuclear Physics, 
630090 Novosibirsk, Russia}

\title{Long-range interactions of hydrogen atoms in excited states. II.\\
Hyperfine-resolved $\bm{2S}$--$\bm{2S}$ system}

\author{U. D. Jentschura}
\affiliation{\addrROLLA}

\author{V. Debierre}
\affiliation{\addrROLLA}

\author{C. M. Adhikari}
\affiliation{\addrROLLA}

\author{A. Matveev}
\affiliation{\addrLEBEDEV}
\affiliation{\addrMUC}

\author{N. Kolachevsky}
\affiliation{\addrLEBEDEV}
\affiliation{\addrMUC}
\affiliation{\addrRQC}

\begin{abstract}
The interaction of two excited hydrogen atoms in metastable states 
constitutes a theoretically interesting 
problem because of the quasi-degenerate $2P_{1/2}$ levels 
which are removed from the $2S$ states only by the Lamb shift.
The total Hamiltonian of the system is 
composed of the van der Waals Hamiltonian, the Lamb shift 
and the hyperfine effects.
The \vdw{} shift becomes commensurate with the 
$2S$--$2P_{3/2}$ fine-structure splitting only for close 
approach ($R<100\,a_0$, where $a_0$ is the Bohr radius) 
and one may thus restrict the discussion to the 
levels with $n=2$ and $J=1/2$ to good approximation.
Because each $S$ or $P$ state splits into an $F=1$ triplet
and an $F=0$ hyperfine singlet (eight states for each atom), the Hamiltonian matrix
{\em a priori} is of dimension $64$. 
A careful analysis of symmetries the problem allows one 
to reduce the dimensionality of the most involved 
irreducible submatrix to $12$.
We determine the Hamiltonian matrices and the 
leading-order van der Waals shifts for states which are
degenerate under the action of the unperturbed Hamiltonian
(Lamb shift plus hyperfine structure).
The leading first- and second-order van der Waals 
shifts lead to interaction energies proportional to $1/R^3$
and $1/R^6$ and are evaluated within the hyperfine 
manifolds. When both atoms are metastable $2S$ states,
we find an interaction energy of order
$E_h \, \chi \, (a_0/R)^6$, where $E_h$ and $\calL$ are the Hartree 
and Lamb shift energies, respectively, and 
$\chi = E_h/\calL \approx 6.22 \times 10^6$ is their ratio.
\end{abstract}


\maketitle


\newpage

%
%
\section{Introduction}
\label{sec1}

Inspired by recent optical measurements of the $2S$ hyperfine splitting using
an atomic beam \cite{KoEtAl2009}, we here aim to carry out an analysis of the
hyperfine-resolved $2S$--$2S$ system composed of two hydrogen atoms.  This paper
follows a previous work of ours (Ref.~\cite{AdEtAl2016vdWi}) in which we analyzed the
long-range interaction between two hydrogen atoms, one of which was in the $1S$
ground state, and the other one in the metastable $2S$ state. Here we turn to the
case where both atoms are in an excited state. For that we use the simplest
case at hand, namely that where both atoms are in the $2S$ state. The
$2S$--$2S$ \vdw{} interaction has been analyzed before in
Refs.~\cite{JoEtAl2002,SiKoHa2011}, but without any reference to the resolution
of the hyperfine splitting \footnote{More accurately, the conclusions of
Ref.~\cite{JoEtAl2002} announce a future work where ``the effects of spin-orbit
coupling and the Lamb shift'' would be taken into account. As far as we could
find, no such work has been published yet.}. The entire problem needs to be
treated using degenerate perturbation theory, because the van der Waals{}
Hamiltonian
couples the reference $2S$ state to neighboring quasi-degenerate $2P$
states. The latter are displaced from the former only by the Lamb shift (in the
case of $2P_{1/2}$) or by the fine structure (in the case of $2P_{3/2}$). As
was noted in Ref.~\cite{AdEtAl2016vdWi}, significant modifications of the long-range
interactions between two atoms result from the presence of quasi-degenerate
states, and the effects lead to observable consequences.  In a more general
context, one may regard our investigations as example cases for a more general
setting, in which two excited atoms interact, while in metastable 
states (with quasi-degenerate levels nearby).

The present work combines the challenges described in Ref.~\cite{JoEtAl2002},
where the $2S$--$2S$ interaction is studied (but without taking account of the
fine and hyperfine structures), with the intricacies of the hyperfine
correction to the long-range interaction of two atoms, which have been studied
in Refs.~\cite{RaLyDa1968a,RaLyDa1968b,RaDa1969,RaIkDa1970}.  Indeed, it had
been anticipated in Ref.~\cite{JoEtAl2002} that a more detailed study of the
combined hyperfine and \vdw{} effects will be required for the $2S$--$2S$
system when a more detailed understanding is sought. The main limitation of the
method followed here is that we will only consider dipole-dipole terms in the
interatomic interaction, in contrast to Refs.~\cite{JoEtAl2002,SiKoHa2011}.
Hence, our analysis only yields reliable results for sufficiently large
interatomic separation. Inspection of the higher-order multipole terms obtained
in Refs.~\cite{JoEtAl2002,SiKoHa2011} clarifies that the dipole-dipole
approximation is already largely valid for interatomic separations of the order
of $R=20\,a_0$. [This is true for the $2S$--$2S$ system, upon which we
focus here. Judging from Fig.~2 in Ref.~\cite{SiKoHa2011}, for higher principal
quantum number ($n=4$), the range of relevance of higher-order multipole terms
extends further out, but these cases are 
beyond the scope of the current investigation.]

Throughout this article, 
we work in SI mksA units and keep all factors of 
$\hbar$ and $c$ in the formulas. 
In the choice of the unit system for this paper, 
we attempt to optimize the accessibility of the 
presentation to two different communities: the 
QED community in general uses the natural 
unit system with $\hbar = c = \epsilon_0 = 1$, 
and the electron mass is denoted as $m$.
The relation $e^2 = 4\pi\alpha$ then allows to 
identify the expansion in the number of quantum 
electrodynamic corrections with powers of the 
fine-structure constant $\alpha$.
This unit system is used, 
e.g., in the investigation reported in Ref.~\cite{Pa2005longrange}
on relativistic corrections to the Casimir--Polder 
interaction (with a strong overlap with QED).
In the atomic unit system, we have 
$|e| = \hbar = m = 1$, and $4 \pi \epsilon_0 = 1$.
The speed of light, in the atomic unit system, is 
$c = 1/\alpha \approx 137.036$. This system of units 
is especially useful for the analysis of purely 
atomic properties without radiative effects.
As the subject of the current study lies in between 
the two mentioned fields of interest, we choose the 
SI mksA unit system as the most appropriate reference frame
for our calculations. The formulas do not become unnecessarily 
complex, and can be evaluated with ease for any 
experimental application.

We organize this paper as follows.  The combination of the orbital and spin
electron angular momenta, and the nuclear spin, add up to give the total
angular momentum of the hydrogen atom; the conserved quantities are discussed
in Sec.~\ref{sec2}, together with the relevant two-atom product wave functions.
In Sec.~\ref{sec3}, we proceed to investigate the Hamiltonian matrices in the
subspaces of the spectrum of the total Hamiltonian into which it naturally
decouples.  Namely, the magnetic projection of the total angular momentum
(summed over both atoms) commutes with the total Hamiltonian, and this leads to
matrix subspaces with $F_z = +2,1,0,-1,-2$. For each one of
these five hyperfine subspaces, we shall identify two irreducible subspaces of
equal dimensionality. This property considerably simplifies the treatment
of the problem.
Finally, some relevant energy differences 
for the $2S$ hyperfine splitting (with the spectator
atom in specific states, namely either $2S$ or $2P$) are analyzed in Sec.~\ref{sec4}.
Conclusions are drawn in Sec.~\ref{conclu}.

%
%
\section{Formalism}
\label{sec2}

%
%
\subsection{Total Hamiltonian of the system}

In order to evaluate the $2S$--$2S$ long-range interaction,
including hyperfine effects,
one needs to diagonalize the Hamiltonian 
\begin{equation}
\label{H}
H = H_{\LS,A} + H_{\LS,B} + 
H_{\HFS,A} + H_{\HFS,B} + H_{\vdW} \,.
\end{equation}
Here, $H_{\rm LS}$ is the Lamb shift Hamiltonian,
while $H_{\rm HFS}$ describes hyperfine effects;
these Hamiltonians have to be added for atoms $A$ and $B$.
They are given as follows,
\begin{subequations}
\begin{align}
\label{HHFS}
H_{\rm HFS}=&\frac{\mu_0}{4\pi}\mu_B\mu_N\,g_sg_p\sum_{i=A,B}\left[\vphantom{\frac{1}{4\pi\left|\vec{r}_i\right|^3}\vec{L}_i\cdot\vec{I}_i}\frac{8\pi}{3}\vec{S}_i\cdot\vec{I}_i\,\delta^3\left(\vec{r}_i\right)\right.\nonumber\\
&\hspace{-25pt}\left.+\frac{3\left(\vec{S}_i\cdot\vec{r}_i\right)\left(\vec{I}_i\cdot\vec{r}_i\right)-\vec{S}_i\cdot\vec{I}_i\,\vec{r}_i^2}{\left|\vec{r}_i\right|^5}+\frac{\vec{L}_i\cdot\vec{I}_i}{\left|\vec{r}_i\right|^3}\right]\\
\label{HLS}
H_{\rm LS}=&\frac{4}{3}\alpha^2\,m c^2\left(\frac{\hbar}{m c}\right)^3\ln\left(\alpha^{-2}\right)\sum_{i=A,B}\delta^3\left(\vec{r}_i\right) \,,
\\[0.0077ex]
\label{vdw}
H_{\rm vdW} =& \;
\alpha \,\hbar c\, \frac{x_A \, x_B + y_A \, y_B - 2 \, z_A \, z_B}{R^3}  \,.
\end{align}
\end{subequations}
Here, $\alpha$ is the fine-structure constant, $m $ the electron mass,
$\vec{r}_i$, $\vec{p}_i$ and $\vec{L}_i$ are the position (relative to
the respective nucleus), linear momentum and orbital angular momentum operators
for electron $i$; also, $\vec{S}_i$ is the spin operator for electron $i$
and $\vec{I}_i$ is the spin operator for proton $i$ [both are
dimensionless]. The electronic and protonic $g$ factors are
$g_s\simeq2.002\,319$ and $g_p\simeq5.585\,695$, while
$\mu_B\simeq9.274\,010\,\times10^{-24}\,\mathrm{A m}^2$ is the Bohr magneton
and $\mu_N\simeq5.050\,784\,\times10^{-27}\,\mathrm{A m}^2$ is the nuclear
magneton.  The subscripts $A$ and $B$ refer to the relative coordinates within
the two atoms, while $R$ is the interatomic distance.  The expression for
$H_{\rm LS}$ shifts $S$ states relative to $P$ states by the Lamb shift, which
is given in the Welton approximation~\cite{ItZu1980}, which is convenient
within the formalism used for the evaluation of matrix elements. 
(The important property of $H_{\rm LS}$ is that it shifts 
$S$ states upward in relation to $P$ states; the prefactor multiplying the 
Dirac-$\delta$ can be adjusted to the observed Lamb shift 
splitting.) Indeed, for
the final calculation of energy shifts, we shall replace
\begin{multline}
\label{defcalL}
\langle 2S_{1/2} | H_{LS} | 2S_{1/2} \rangle -
\langle 2P_{1/2} | H_{LS} | 2P_{1/2} \rangle \\
=\frac{4\alpha}{3 \pi}\, \frac{\alpha^4}{8} \, m \,c^2 \, \ln( \alpha^{-2}) 
\to \calL \,,
\end{multline}
where $\calL = h \, 1057.845(9) \, {\rm MHz}$ is the 
``classic'' $2S$--$2P_{1/2}$ Lamb shift~\cite{LuPi1981}.
The Hamiltonian $H$ given in Eq.~\eqref{H} 
defines the zero of the energy to be the 
hyperfine centroid frequency of the $2P_{1/2}$ states.
The result for $H_{\rm HFS}$ in the given form
is taken from Ref.~\cite{JeYe2006}.
The Hamiltonians $H_{\HFS,A}$ and $H_{\HFS,B}$ are
obtained from $H_\HFS$ by specializing the coordinate
$\vec r$ to be the relative coordinate (electron-proton) 
in atoms $A$ and $B$, respectively,
and correspondingly for $H_{\LS,A}$ and $H_{\LS,B}$.

We shall focus on the interatomic separation regime where the \vdw{} energy
is commensurate with the hyperfine splitting and 
Lamb shift energies, but much smaller than the 
fine structure (the $2P_{1/2}$--$2P_{3/2}$ splitting
and likewise, the $2S$--$2P_{3/2}$ splitting).
Hence,
\begin{equation}
E_{\vdW} \sim E_{\HFS} \sim \calL \ll E_{\FS} \,.
\end{equation}
This is fulfilled for $R > 100 \, a_0$, as can be seen from 
Eq.~\eqref{vdw} and will be confirmed later. 
Hence, we only consider $2S$ and $2P_{1/2}$ states.
We shall neglect the influence of the 
$2P_{3/2}$ states, assuming that they are sufficiently 
displaced.
Because the \vdw{} interaction~\eqref{vdw} has nonvanishing
diagonal elements between $2S$ and $2P$ states,
the interaction energy between the two $2S$ atoms can be of order $1/R^3$.

The $z$ component of the total angular momentum operator
of both atoms is
\begin{equation}  \label{eq:BothSpins}
\begin{aligned} [b]
F_z &= F_{z,A}+F_{z,B} =
J_{z,A}+J_{z,B} +
I_{z,A}+I_{z,B}\\
&=L_{z,A} + L_{z,B} 
+ S_{z,A}
+ S_{z,B}
+ I_{z,A}
+ I_{z,B}\\
&= L_{z,A} + L_{z,B} 
+ \tfrac12 \, \sigma_{e,z,A}
+ \tfrac12 \, \sigma_{e,z,B}\\
&\hspace{66.25pt}
+ \tfrac12 \, \sigma_{p,z,A}
+ \tfrac12 \, \sigma_{p,z,B} \,,
\end{aligned}
\end{equation}
where $\vec J = \vec L + \vec S$ is the total
angular momentum of the electron.
Let us investigate if $F_z$
commutes with the total Hamiltonian $H$.
In Eq.~\eqref{eq:BothSpins}, the subscript $e$ denotes the electron,
while $p$ denotes the proton. The 
following commutators vanish separately,
$\left[S_{z,a}+S_{z,b},H_{\mathrm{LS}}\right]=
\left[S_{z,a}+S_{z,b},H_{\mathrm{vdW}}\right]=
\left[I_{z,a}+I_{z,b},H_{\mathrm{LS}}\right]= 
\left[I_{z,a}+I_{z,b},H_{\mathrm{vdW}}\right]=0$.
We then turn to the non-trivial commutators. For that, it is very useful to
notice that the orbital angular momentum $\vec{L}_i$ of electron $i$
commutes with all spherically symmetric functions of the radial position
operator $|\vec{r}_i|$ of the same electron. This immediately yields
$\left[L_{z,a}+L_{z,b},H_{\mathrm{LS}}\right]=0$.
We can also show that
\begin{multline} 
\left[S_{z,a}+S_{z,b},H_{\mathrm{HFS}}\right]+
\left[ I_{z,a}+ I_{z,b},H_{\mathrm{HFS}}\right]\\
+\left[L_{z,a}+L_{z,b},H_{\mathrm{HFS}}\right]=0 \,, \\
\left[L_{z,a}+L_{z,b},H_{\mathrm{vdW}}\right] =
\alpha\,\hbar c\,\frac{\mathrm{i}\hbar}{R^3}
\left[y_A\,x_B+x_A\,y_B \right.  \\
\left. -y_A\,x_B-x_A\,y_B \right] = 0.
\end{multline}
The component $F_z$ of the total angular
momentum of the two-atom system 
[see Eq.~(\ref{eq:BothSpins})]
thus commutes with the total Hamiltonian $H$. We
can classify states according to the
eigenvalues of the operator $F_z= F_{z,a}+F_{z,b}$.

%
%
\subsection{Addition of Momenta and Total Hyperfine Quantum Number}

In order to calculate the matrix elements of the total Hamiltonian (\ref{H}), 
we first need to identify the relevant states of the two atoms.
For each atom, we easily identify the following 
quantum numbers within the hyperfine manifolds:
\begin{subequations}
\begin{align}
2S_{1/2}(F=0): \ell = 0, \, J =\tfrac12, \, F=0 \; \Rightarrow \; g_F=1 \,, \\
2S_{1/2}(F=1): \ell = 0, \, J =\tfrac12, \, F=1 \; \Rightarrow \; g_F=3 \,, \\
2P_{1/2}(F=0): \ell = 1, \, J =\tfrac12, \, F=0 \; \Rightarrow \; g_F=1 \,, \\
2P_{1/2}(F=1): \ell = 1, \, J =\tfrac12, \, F=1 \; \Rightarrow \; g_F=3 \,.
\end{align}
\end{subequations}
Here $\ell$, $J$, and $F$ are the electronic orbital angular momentum, the
total (orbital$+$spin) electronic angular momentum and the total
(electronic$+$protonic) atomic angular momentum, while $g_F = 2F+1$ is the
number of states. At this stage, 
we remember that we discarded $2P_{3/2}$ states from our
treatment because of their relatively large energy separation from $2S_{1/2}$
and $2P_{1/2}$ states.  Thus, we have a total of eight states per atom. For the
system of two atoms, we have $8\times8=64$ states.  Due to the conservation of
the total hyperfine quantum number $F_{z}=F_{z,a}+$ $F_{z,b}$, established
above, the $64$-dimensional Hilbert space is decomposed into five subspaces as 
\begin{subequations}
\label{FzSeparation}
\begin{align}
F_z = F_{z,a}+F_{z,b} = \pm 2 & \; \Rightarrow g=4 \,,
\\
F_z = F_{z,a}+F_{z,b} = \pm 1 & \; \Rightarrow g=16 \,,
\\
F_z = F_{z,a}+F_{z,b} = 0     & \; \Rightarrow g=24 \,.
\end{align}
\end{subequations}
The most complicated case is the subspace for which $F_z = 0$,
in which case the Hamiltonian matrix is, \emph{a priori}, $24$-dimensional. 
Thus, we have to
generate the matrix, diagonalize it and choose the eigenvalues which
corresponds to the unperturbed (with respect to dipole-dipole interaction)
states. 


Let us add angular momenta to obtain the single-atom states of definite
hyperfine quantum number. First, we add the electron spin with its orbital
angular momentum to obtain the $J=1/2$ states within the $n=2$ manifold of
hydrogen. These are given as follows,
\begin{subequations}
\begin{align}
\left\vert \ell=0,J_{z}=\pm \tfrac12 \right\rangle 
=& \; \left\vert \pm\right\rangle_e \; 
\left\vert \ell=0,m=0\right\rangle_e
= \left\vert \pm\right\rangle_e \; 
\left\vert 0, 0\right\rangle_e \,, 
\\
\left\vert \ell=1,J_{z}=\pm \tfrac12 \right\rangle  =& \;
\mp\left[\frac{1}{\sqrt{3}} \; \vphantom{\sqrt{\frac{2}{3}}}
\left\vert \pm\right\rangle_e \; 
\left\vert 1,0\right\rangle_e \right. \nonumber\\
&\hspace{25pt}\left.-
\sqrt{\frac{2}{3}} \;
\left\vert \mp\right\rangle_e \;
\left\vert 1,\pm1\right\rangle_e\right] \,.
\end{align}
\end{subequations}
Here, $\left\vert \pm\right\rangle_e$ is the electron spin state, and
$|\ell,m\rangle_e$ denotes the Schr\"{o}dinger eigenstate (without spin). 
The principal quantum is $n=2$ throughout.
We also remember that the 
$J=3/2$ states are displaced by the fine structure shift and, therefore,
far away in the energy landscape given the scale of energies considered here.
With the help of Clebsch--Gordan coefficients,
we add the nuclear (proton) spin $| \pm \rangle_p$ 
to obtain the eight states in the
single-atom hyperfine basis.  First, we have for the four $S$ states,
\begin{widetext}
\begin{subequations} \label{eq:2SHyp}
\begin{align}
\left\vert \ell = 0,F=0,F_{z}=0\right\rangle =& \;
-\frac{
\left\vert +\right\rangle_p \,
\left\vert -\right\rangle_e -
\left\vert -\right\rangle_p \,
\left\vert +\right\rangle_e}{\sqrt{2}} \;
\left\vert 0,0\right\rangle_e \,, \\
\left\vert \ell = 0,F=1,F_{z}=0\right\rangle =& \;
\frac{
\left\vert + \right\rangle_p \,
\left\vert -\right\rangle_e +
\left\vert -\right\rangle_p \,
\left\vert +\right\rangle_e}{\sqrt{2}}\left\vert 0,0\right\rangle_e 
\,, \\
\left\vert \ell = 0,F=1,F_{z}=\pm1\right\rangle =& \;
\left\vert \pm\right\rangle_p \,
\left\vert \pm\right\rangle_e \,
\left\vert 0,0\right\rangle_e \,.
\end{align}
\end{subequations}
The $P$ states are more complicated,
\begin{subequations} \label{eq:2PHyp}
\begin{align}
\left\vert \ell=1,F=0,F_{z}=0\right\rangle =& \;
\frac{1}{\sqrt{3}} 
\left\vert +\right\rangle_p 
\left\vert +\right\rangle_e 
\left\vert 1,-1\right\rangle_e 
- \frac{1}{\sqrt{6}} 
\left\vert +\right\rangle_p 
\left\vert -\right\rangle_e 
\left\vert 1,0\right\rangle_e +
\frac{1}{\sqrt{3}} 
\left\vert -\right\rangle_p 
\left\vert -\right\rangle_e
\left\vert 1,1\right\rangle_e 
- \frac{1}{\sqrt{6}} 
\left\vert -\right\rangle_p 
\left\vert +\right\rangle_e
\left\vert 1,0\right\rangle_e
\\
\left\vert \ell=1,F=1,F_{z}=0\right\rangle =& \;
-\frac{1}{\sqrt{3}}
\left\vert +\right\rangle_p 
\left\vert +\right\rangle_e
\left\vert 1,-1\right\rangle_e +
\frac{1}{\sqrt{6}}
\left\vert +\right\rangle_p 
\left\vert -\right\rangle_e
\left\vert 1,0\right\rangle_e +
\frac{1}{\sqrt{3}}
\left\vert -\right\rangle_p 
\left\vert -\right\rangle_e 
\left\vert 1,1\right\rangle_e -
\frac{1}{\sqrt{6}} 
\left\vert -\right\rangle_p
\left\vert +\right\rangle_e
\left\vert 1,0\right\rangle_e
\\
\left\vert \ell=1,F=1,F_{z}=\pm1\right\rangle =& \;
\mp\frac{1}{\sqrt{3}} 
\left\vert \pm\right\rangle_p \,
\left[ 
\left\vert \pm\right\rangle_e 
\left\vert 1,0\right\rangle_e -
\sqrt{2}
\left\vert \mp\right\rangle_e \,
\left\vert 1,\pm1\right\rangle_e \right] \,.
\end{align}
\end{subequations}
\end{widetext}
In the following, we shall use the notation 
$| \ell, F, F_z \rangle$ for the eigenstates of the 
unperturbed Hamiltonian 
\begin{equation}
H_0 = H_{\HFS,A} + H_{\HFS,B}
+ H_{\LS,A} + H_{\LS,B}  \,,
\end{equation}
within the $2S$--$2P_{1/2}$ manifold.
The notation $| \ell, F, F_z \rangle$ is rather 
intuitive; the first entry clarifies if we have an 
$S$ (with $\ell = 0$) or a $P$ state (with $\ell = 1$),
the second entry specifies if we have a
hyperfine triplet ($F=1$) or a 
hyperfine singlet ($F=0$) state,
and the last entry is the magnetic projection
of the total angular momentum.

%
%
\subsection{Matrix Elements of the Total Hamiltonian}

We now turn to the computation of the matrix elements of the total Hamiltonian
(\ref{H}) in the space spanned by the two-atom states which are product states
built from any two states of the types~(\ref{eq:2SHyp}) and~(\ref{eq:2PHyp}).
We choose a basis in which the Lamb shift and hyperfine Hamiltonians are
diagonal, so that the only non-trivial task is to determine the matrix elements
of the van der Waals interaction Hamiltonian.

With the definition of the spherical unit vectors \cite{VaMoKh1988},
\begin{subequations}
\begin{align}
\hat e_+ =& \; -\frac{1}{\sqrt{2}} \, (\hat e_x + \ii \, \hat e_y) \,,
\\
\hat e_- =& \; \frac{1}{\sqrt{2}} \, (\hat e_x - \ii \, \hat e_y) \,,
\\[0.0077ex]
\hat e_0 =&\; \hat e_z \,,
\end{align}
\end{subequations}
and the states defined by (\ref{eq:2SHyp}) and (\ref{eq:2PHyp}), we obtain the
non-zero matrix elements of the electronic position operator $\vec{r}$ as
follows:
\begin{subequations}
\label{django}
\begin{align}
\left\langle 0,0,0\right\vert \vec{r}\left\vert 1,1,0\right\rangle  &
=\sqrt{3} \, a_0 \, \hat e_{z} \,, \\
\left\langle 0,0,0\right\vert \vec{r}\left\vert 1,1,\pm1\right\rangle  &
=\sqrt{3} \, a_0 \, \hat e_{\pm} \,, \\
\left\langle 0,1,0\right\vert \vec{r}\left\vert 1,0,0\right\rangle  &
=\sqrt{3} \, a_0 \, \hat e_{z} \,, \\
\left\langle 0,1,\pm 1\right\vert \vec{r}\left\vert 1,0,0 \right\rangle  &
=\sqrt{3} \, a_0 \left( \hat e_\pm \right)^* \,, \\
\left\langle 0,1, \pm 1\right\vert \vec{r}\left\vert 1,1,\pm1\right\rangle  &
= \pm \sqrt{3} \, a_0 \, \hat e_{z} \,,\\
\left\langle 0,1,\pm 1\right\vert \vec{r}\left\vert 1,1,0\right\rangle  &
= \pm \sqrt{3} \, a_0 \, \hat e_\mp \,, \\
\left\langle 0,1,0\right\vert \vec{r}\left\vert 1,1,\pm 1\right\rangle  &
= \mp \sqrt{3} \, a_0 \, \hat e_\pm \,.
\end{align}
\end{subequations}
All the other matrix elements vanish. We define the parameters
\begin{subequations} 
\label{parameters}
\begin{align}
\calH &\equiv 
\frac{\alpha^4}{18} \, g_N \, \frac{m }{m_p}\,m \,c^2 
\to h \, 59.1856114(22) \, {\rm MHz}
\,,\label{eq:HFSplit}\\
\calL &\equiv \frac{\alpha^5}{6 \, \pi} \, 
\ln(\alpha^{-2}) \, m  \, c^2 
\to h \, 1057.845(9) \, {\rm MHz}
\,, \label{eq:LSSplit}\\
\calV &\equiv 3\,\alpha\,\hbar c\,\frac{a_0^2}{R^3}\, 
\label{eq:Inter} \,,
\end{align}
\end{subequations}
where the data used after the replacements
indicates one-third of the hyperfine splitting of 
the $2S$ state~\cite{KoEtAl2009} and 
the classic Lamb shift~\cite{LuPi1981}, respectively.
These data are used in all figures for the 
plots of the distance-dependent energy levels.
Note that $\calH$ and $\calL$ obviously are constants, whereas
$\calV$ depends on the interatomic separation $R$. The expectation values
of the hyperfine $H_\HFS$ and Lamb shift $H_\LS$ Hamiltonians (for states of
both atoms $A$ and $B$) are given as follows
\begin{subequations}
\begin{align}
\langle \ell, F, M_F | H_\LS| \ell, F, M_F \rangle =& \; \calL\, \delta_{\ell 0},
\\
\langle 0, 1, M_F | H_\HFS | 0, 1, M_F \rangle =& \; \frac34 \, \calH \,,
\\
\langle 0, 0, 0 | H_\HFS | 0, 0, 0 \rangle =& \; -\frac94 \, \calH \,,
\\
\langle 1, 1, M_F | H_\HFS | 1, 1, M_F \rangle =& \; \frac14 \, \calH \,,
\\
\langle 1, 0, 0 | H_\HFS | 1, 0, 0 \rangle =& \; -\frac34 \, \calH \,.
\end{align}
\end{subequations}
The hyperfine splitting energy between $2P_{1/2}(F=1)$ and 
$2P_{1/2}(F=0)$ states thus amounts to $\calH$, 
while the $S$-state splitting is $3 \calH$.
Additionally, the energies of the $S$ states are lifted upward by 
$\calL$, irrespective of the hyperfine effects.
For the product state of atoms $A$ and $B$,
we shall use the notation
\begin{equation}
| (\ell_A, F_A, F_{z,A})_A \, (\ell_B, F_B, F_{z,B})_B \, \rangle \,,
\end{equation}
which summarizes the quantum numbers of both atoms.
We anticipate that some of the
eigenstates of the combined and total
Hamiltonian (Lamb shift plus hyperfine effects plus \vdw{})
do not decouple into simple unperturbed eigenstates of the form
$| (\ell_A, F_A, F_{z,A})_A \, (\ell_B, F_B, F_{z,B})_B \rangle$
but may require the use of superpositions of these states,
as we had already experienced for the $(1S;2S)$ interaction
in Ref.~\cite{AdEtAl2016vdWi}.

\begin{figure*}[t!]
\begin{center}
\begin{minipage}{0.91\linewidth}
\begin{center}
\includegraphics[width=0.8\linewidth]{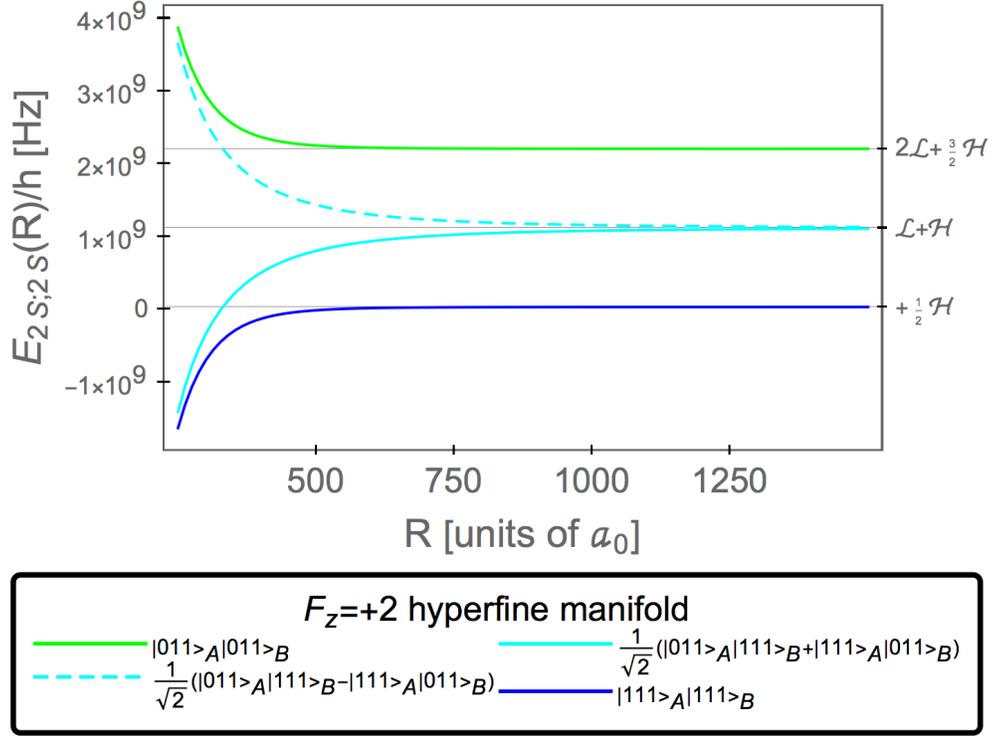}
\caption{(Color online.) Evolution of the energy levels 
within the $F_z=+2$ hyperfine manifold
as a function of interatomic separation. The eigenstates given in the legend
are only asymptotic; for finite separation these states mix.
One has $\calH = 0.055\,949 \, \calL$ according to 
Eq.~\eqref{parameters}. The unit of energy used for the 
ordinate axis is interaction energy 
divided by the Planck constant $h$ (left ordinate axis)
and given in Hertz (Hz).
On the right ordinate axis, we use
the Lamb shift $\calL$ as defined in Eq.~\eqref{defcalL}
as an alternative unit of frequency. The Born--Oppenheimer approximation is 
used in plotting the interaction energy as a function of the 
internuclear distance $R$.
\label{fig1}}
\end{center}
\end{minipage}
\end{center}
\end{figure*}

%
%
\section{Hamiltonian Matrices in the Hyperfine Subspaces}
\label{sec3}

%
%
\subsection{Manifold $\maybebm{F_z=+2}$}
\label{sec31}

We have already pointed out that the $n=2$, $J=1/2$ Hilbert space
naturally separates 
into subspaces with fixed total hyperfine quantum number $F_z=F_{z,a}+F_{z,b}$,
according to Eq.~\eqref{FzSeparation}.
We can identify two irreducible subspaces within the $F_z=+2$ manifold: 
the subspace $\mathrm{\Rmnum{1}}$ is composed of the 
states
\begin{subequations}
\begin{align}
| \phi_{1}^{\left(\mathrm{\Rmnum{1}}\right)} \rangle &= 
| (0,1,1)_A \, (0,1,1)_B \rangle \,,\\
| \phi_{2}^{\left(\mathrm{\Rmnum{1}}\right)} \rangle &= 
| (1,1,1)_A \, (1,1,1)_B \rangle \,,
\end{align}
\end{subequations}
where the Hamiltonian matrix reads
\begin{equation}
\label{matFz2Irr1}
H_{F_z=+2}^{\left(\mathrm{\Rmnum{1}}\right)} = \left(
\begin{array}{cc}
 2 \calL+\tfrac32 \calH & -2 \calV \\
 -2 \calV & \tfrac12 \calH \\
\end{array}
\right) \,.
\end{equation}
Subspace $\mathrm{\Rmnum{2}}$ is composed of the states
\begin{subequations}
\begin{align}
| \phi_{1}^{\left(\mathrm{\Rmnum{2}}\right)} \rangle =& \;
| (0,1,1)_A \, (1,1,1)_B \rangle \,,\\
| \phi_{2}^{\left(\mathrm{\Rmnum{2}}\right)} \rangle =& \;
| (1,1,1)_A \, (0,1,1)_B \rangle \,,
\end{align}
\end{subequations}
where the Hamiltonian matrix reads
\begin{equation}
\label{matFz2Irr2}
H_{F_z=+2}^{\left(\mathrm{\Rmnum{2}}\right)} = \left(
\begin{array}{cccc}
 \calL+\calH & -2 \calV \\
 -2 \calV & \calL+\calH \\
\end{array}
\right) \,.
\end{equation}
These subspaces are completely uncoupled. Namely, no state in subspace
$\mathrm{\Rmnum{1}}$ is coupled to a state in subspace $\mathrm{\Rmnum{2}}$.

The eigenvalues of $H_{F_z=+2}^{\left(\mathrm{\Rmnum{1}}\right)}$ are given by
\begin{subequations} \label{eq:EV2PM1}
\begin{align}
E_+^{\left(\mathrm{\Rmnum{1}}\right)} =& \; \calH + \calL + 
\sqrt{4 \calV^2 + (\tfrac12 \calH + \calL)^2 } 
\nonumber\\[0.0077ex]
=& \; \tfrac32 \, \calH + 2 \calL + 
4 \, \frac{\calV^2}{\calH +2 \calL} 
+ \calO(\calV^4) \,,
\\[0.0077ex]
E_-^{\left(\mathrm{\Rmnum{1}}\right)} =& \; \calH + \calL - 
\sqrt{4 \calV^2 + (\tfrac12 \calH + \calL)^2 } 
\nonumber\\[0.0077ex]
=& \; \tfrac12 \, \calH - 4 \, \frac{\calV^2}{\calH +2 \calL} 
+ \calO(\calV^4) \,,
\end{align}
\end{subequations}
with the corresponding eigenvectors
\begin{subequations} 
\begin{align}
| u_+^{\left(\mathrm{\Rmnum{1}}\right)} \rangle &=
\frac{1}{\sqrt{a^2+b^2}}\left( a \, | \phi_{1}^{\left(\mathrm{\Rmnum{1}}\right)} \rangle + 
b \, | \phi_{2}^{\left(\mathrm{\Rmnum{1}}\right)} \rangle\right) \,,\\
| u_-^{\left(\mathrm{\Rmnum{1}}\right)} \rangle &=
\frac{1}{\sqrt{a^2+b^2}}\left(b \, | \phi_{1}^{\left(\mathrm{\Rmnum{1}}\right)} \rangle - 
a \, | \phi_{2}^{\left(\mathrm{\Rmnum{1}}\right)} \rangle\right) \,.
\end{align}
\end{subequations}
Here the coefficients $a$ and $b$ are given by
\begin{subequations} \label{eq:UglyCoeff}
\begin{align}
a&=-\frac{2\calL+\calH+\sqrt{\left(2\calL+\calH\right)^2+
\left(4\calV\right)^2}}{4\calV}\,,\\ b&=1\,.
\end{align}
\end{subequations}
The eigenenergies of $H_{F_z=+2}^{\left(\mathrm{\Rmnum{2}}\right)}$ are given by
\begin{equation} 
\label{eq:EV2PM2}
E_\pm^{\left(\mathrm{\Rmnum{2}}\right)} = 
\calH \pm \calL \pm 2 \calV  \,,
\end{equation}
with the corresponding eigenvectors,
\begin{equation}
| u_\pm^{\left(\mathrm{\Rmnum{2}}\right)} \rangle = 
\frac{1}{\sqrt{2}} \, ( | \phi_{1}^{\left(\mathrm{\Rmnum{2}}\right)} \rangle \pm 
| \phi_{2}^{\left(\mathrm{\Rmnum{2}}\right)} \rangle ) \,.
\end{equation}
For $\calV \to 0$, which corresponds to the large separation limit
$R\to+\infty$, these eigenvalues tend toward the (degenerate) diagonal entries
of the matrix $H_{F_z=+2}^{\left(\mathrm{\Rmnum{2}}\right)}$.

The eigenstates within the degenerate subspace $\mathrm{\Rmnum{2}}$ 
experience a shift of first order in the van der Waals interaction energy $\calV$,
because of the degeneracy of the diagonal entries $\calL + \calH$
in Eq.~\eqref{matFz2Irr2}; this pattern will be observed for other 
subspaces in the following.
In Fig.~\ref{fig1}, we plot the evolution of the eigenvalues
(\ref{eq:EV2PM1}) and (\ref{eq:EV2PM2}) with respect to interatomic separation.
The two levels within the subspace $\mathrm{\Rmnum{2}}$ noticeably
experience a far larger interatomic interaction shift
from their asymptotic value $\calL+\calH$,
commensurate with the parametric estimate of the corresponding energy shifts.

\begin{figure*}[t!]
\begin{center}
\begin{minipage}{0.91\linewidth}
\begin{center}
\includegraphics[width=0.8\linewidth]{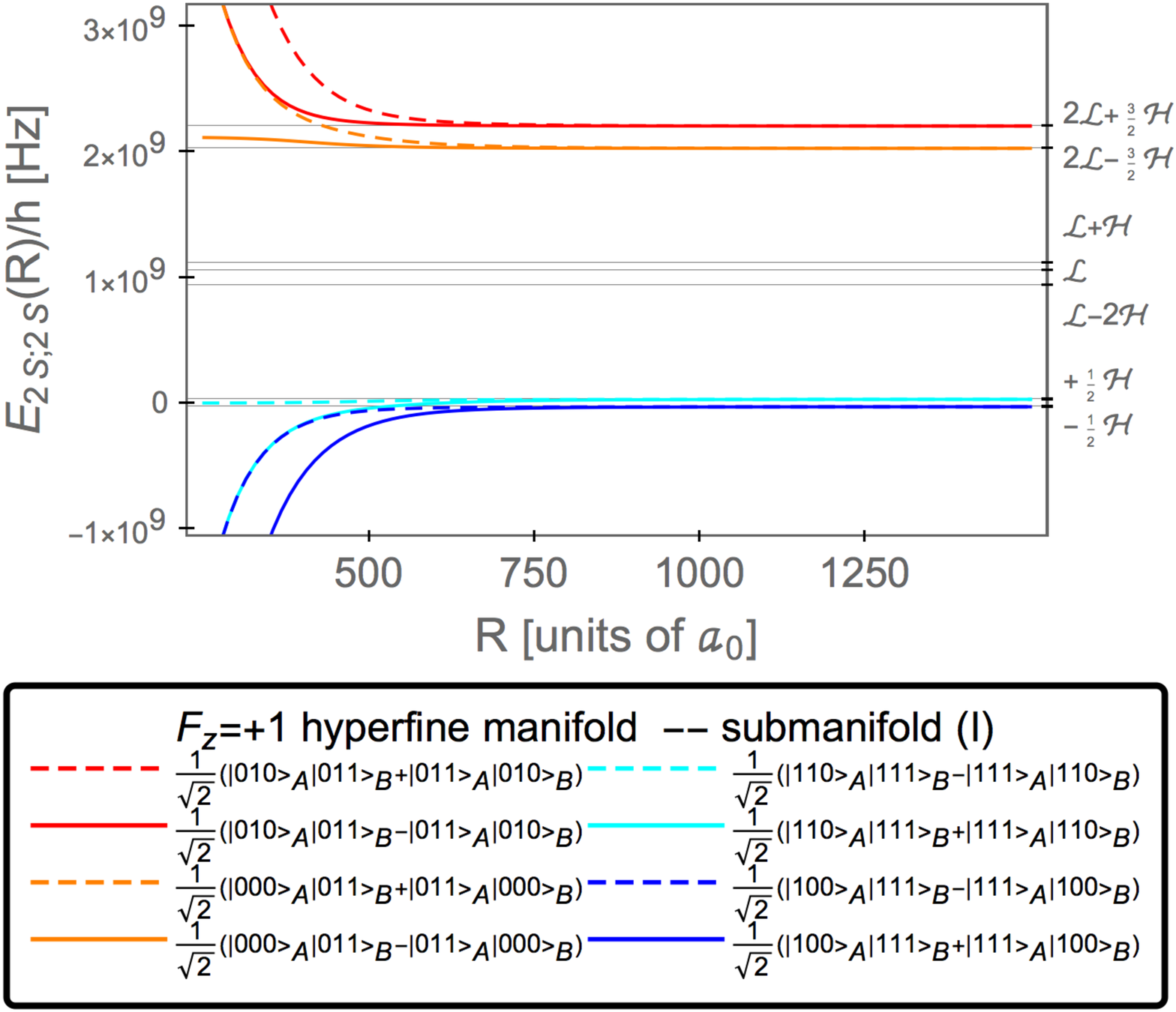}
\caption{(Color online.) Evolution of the 
$S$--$S$ and $P$--$P$ energy levels of the submanifold $\mathrm{\Rmnum{1}}$
within the $F_z=+1$ hyperfine manifold as a function of interatomic separation.
The asymptotic eigenstates given in the legend mix 
for finite separation. The labeling of the axes is as in Fig.~\ref{fig1}.
\label{fig2}}
\end{center}
\end{minipage}
\end{center}
\end{figure*}

\begin{figure*}[t!]
\begin{center}
\begin{minipage}{0.91\linewidth}
\begin{center}
\includegraphics[width=0.7\textwidth]{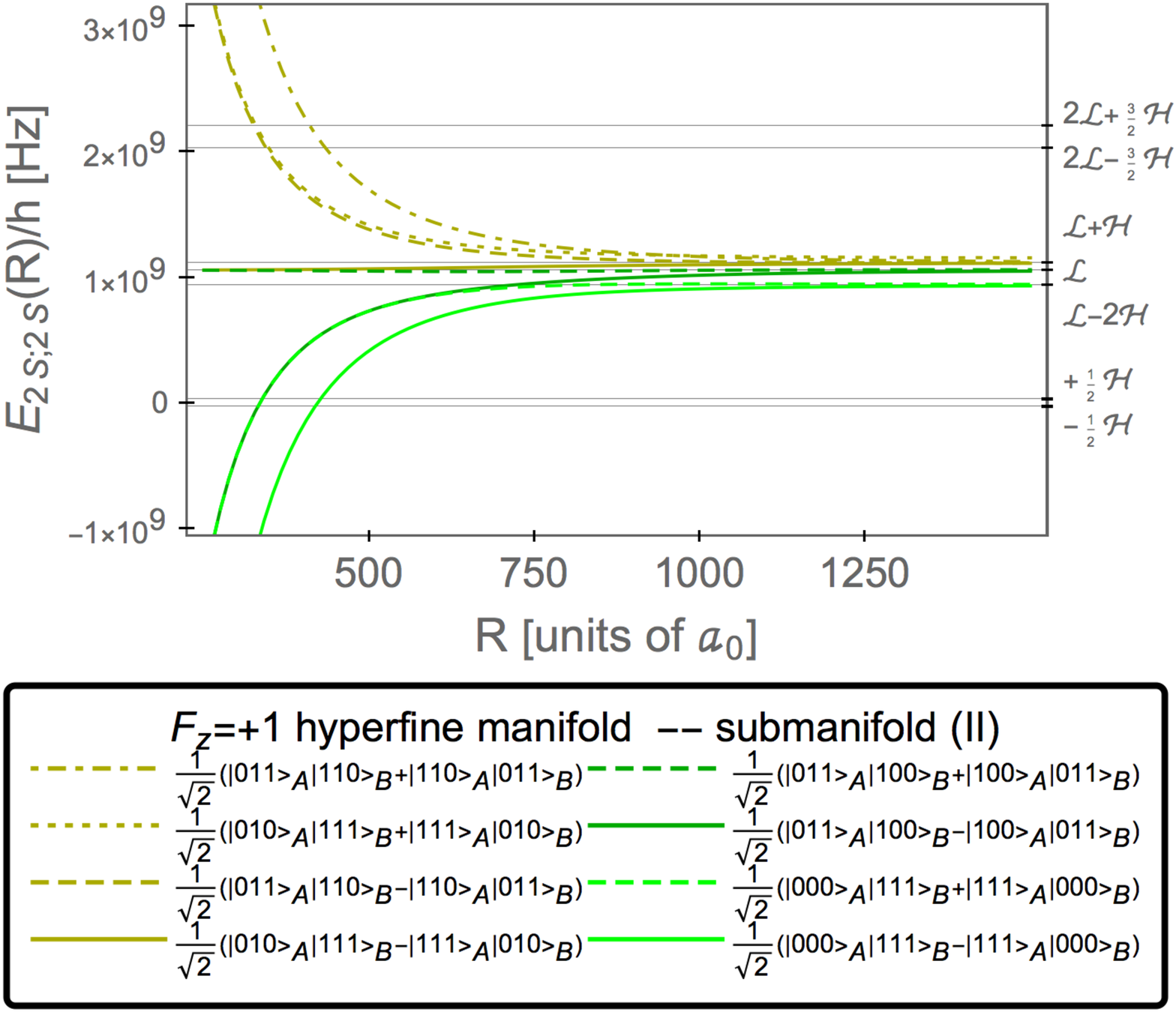}
\caption{(Color online.) Evolution of the energy 
levels of the submanifold $\mathrm{\Rmnum{2}}$
within the $F_z=+1$ hyperfine manifold as a function of interatomic separation.
The eigenstates given in the legend are only asymptotic.
The curve for the seventh state in the legend (counted from the top)
has been slightly offset for better readability, in actuality it is virtually
indistinguishable from that for the sixth state in the legend.
\label{fig3}}
\end{center}
\end{minipage}
\end{center}
\end{figure*}

%
%
\subsection{Manifold $\maybebm{F_z=+1}$}
\label{sec32}

We can identify two irreducible subspaces within the $F_z=+1$ manifold.
Subspace $\mathrm{\Rmnum{1}}$ is composed of the 
following states, with both atoms either being in $S$,
or both in $P$ states,
\begin{widetext}
\begin{equation} 
\label{statesFz1Irr1}
\begin{aligned} 
| \psi_{1}^{\left(\mathrm{\Rmnum{1}}\right)} \rangle =& \;
| (0,0,0)_A \, (0,1,1)_B \rangle \,, \qquad
| \psi_{2}^{\left(\mathrm{\Rmnum{1}}\right)} \rangle =
| (0,1,0)_A \, (0,1,1)_B \rangle \,, \qquad
| \psi_{3}^{\left(\mathrm{\Rmnum{1}}\right)} \rangle =
| (0,1,1)_A \, (0,0,0)_B \rangle \,,
\\[0.0077ex]
| \psi_{4}^{\left(\mathrm{\Rmnum{1}}\right)} \rangle =& \;
| (0,1,1)_A \, (0,1,0)_B \rangle \,, \qquad
| \psi_{5}^{\left(\mathrm{\Rmnum{1}}\right)} \rangle =
| (1,0,0)_A \, (1,1,1)_B \rangle \,, \qquad
| \psi_{6}^{\left(\mathrm{\Rmnum{1}}\right)} \rangle =
| (1,1,0)_A \, (1,1,1)_B \rangle \,,
\\[0.0077ex]
| \psi_{7}^{\left(\mathrm{\Rmnum{1}}\right)} \rangle =& \;
| (1,1,1)_A \, (1,0,0)_B \rangle \,, \qquad
| \psi_{8}^{\left(\mathrm{\Rmnum{1}}\right)} \rangle =
| (1,1,1)_A \, (1,1,0)_B \rangle \,,
\end{aligned}
\end{equation} 
and the Hamiltonian matrix reads
\begin{align}
\label{matFz1Irr1}
H_{F_z=+1}^{\left(\mathrm{\Rmnum{1}}\right)} =& \; \left(
\begin{array}{cccccccc}
 2 \calL-\tfrac32 \calH & 0 & 0 & 0 & 0 & -2 \calV & \calV & - \calV \\
 0 & 2 \calL+\tfrac32 \calH & 0 & 0 & -2 \calV & 0 & - \calV & \calV \\
 0 & 0 & 2 \calL-\tfrac32 \calH & 0 & \calV & - \calV & 0 & -2 \calV \\
 0 & 0 & 0 & 2 \calL+\tfrac32 \calH & - \calV & \calV & -2 \calV & 0 \\
 0 & -2 \calV & \calV & - \calV & -\tfrac12 \calH & 0 & 0 & 0 \\
 -2 \calV & 0 & - \calV & \calV & 0 & \tfrac12 \calH & 0 & 0 \\
 \calV & - \calV & 0 & -2 \calV & 0 & 0 & -\tfrac12 \calH & 0 \\
 - \calV & \mathcal {V} & -2 \calV & 0 & 0 & 0 & 0 & \tfrac12 \calH
\end{array}
\right) \,.
\end{align}
Subspace $\mathrm{\Rmnum{2}}$ is composed of the following states,
where one atom is in a $S$, and the other, in a $P$ state,
\begin{equation}
\label{statesFz1Irr2}
\begin{aligned} [b]
| \psi_{1}^{\left(\mathrm{\Rmnum{2}}\right)} \rangle =& \;
| (0,0,0)_A \, (1,1,1)_B \rangle \,, \qquad
| \psi_{2}^{\left(\mathrm{\Rmnum{2}}\right)} \rangle = 
| (0,1,0)_A \, (1,1,1)_B \rangle \,, \qquad
| \psi_{3}^{\left(\mathrm{\Rmnum{2}}\right)} \rangle = 
| (0,1,1)_A \, (1,0,0)_B \rangle \,,
\\[0.0077ex]
| \psi_{4}^{\left(\mathrm{\Rmnum{2}}\right)} \rangle =& \;
| (0,1,1)_A \, (1,1,0)_B \rangle \,, \qquad
| \psi_{5}^{\left(\mathrm{\Rmnum{2}}\right)} \rangle = 
| (1,0,0)_A \, (0,1,1)_B \rangle \,, \qquad
| \psi_{6}^{\left(\mathrm{\Rmnum{2}}\right)} \rangle = 
| (1,1,0)_A \, (0,1,1)_B \rangle \,,
\\[0.0077ex]
| \psi_{7}^{\left(\mathrm{\Rmnum{2}}\right)} \rangle =& \; 
| (1,1,1)_A \, (0,0,0)_B \rangle \,, \qquad
| \psi_{8}^{\left(\mathrm{\Rmnum{2}}\right)} \rangle = 
| (1,1,1)_A \, (0,1,0)_B \rangle \,,
\end{aligned}
\end{equation}
and the Hamiltonian matrix reads
\begin{align}
\label{matFz1Irr2}
H_{F_z=+1}^{\left(\mathrm{\Rmnum{2}}\right)} =& \; \left(
\begin{array}{cccccccc}
 \calL-2 \calH & 0 & 0 & 0 & 0 & -2 \calV & \calV & - \calV \\
 0 & \calL+\calH & 0 & 0 & -2 \calV & 0 & - \calV & \calV \\
 0 & 0 & \calL & 0 & \calV & - \calV & 0 & -2 \calV \\
 0 & 0 & 0 & \calL+\calH & - \calV & \calV & -2 \calV & 0 \\
 0 & -2 \calV & \calV & - \calV & \calL & 0 & 0 & 0 \\
 -2 \calV & 0 & - \calV & \calV & 0 & \calL+\calH & 0 & 0 \\
 \calV & - \calV & 0 & -2 \calV & 0 & 0 & \calL-2 \calH & 0 \\
 - \calV & \mathcal {V} & -2 \calV & 0 & 0 & 0 & 0 & \calL+\calH
\end{array}
\right) \,.
\end{align}
\end{widetext}
These two submanifolds are, again, completely uncoupled,
as a consequence of the selection rules between $S$ and $P$ states.
One observes that within the subspace $\mathrm{\Rmnum{1}}$, 
no two degenerate levels are coupled
to each other, resulting in second-order van der Waals energy 
shifts. On the other hand, the following subspaces, within the subspace
$\mathrm{\Rmnum{2}}$, can be identified as being degenerate with respect to the
unperturbed Hamiltonian, and having states coupled by nonvanishing off-diagonal
elements. We first have a subspace spanned by
\begin{equation}
\label{IMPORTANT}
| \psi^{\left(A\right)}_{1} \rangle = 
| \psi^{\left(\mathrm{\Rmnum{2}}\right)}_{1} \rangle \,,
\qquad
| \psi^{\left(A\right)}_{2} \rangle = 
| \psi^{\left(\mathrm{\Rmnum{2}}\right)}_{7} \rangle \,.
\end{equation}
These states are composed of a singlet $S$ and a triplet 
$P$ state, and hence the  diagonal entries in the 
Hamiltonian matrix are 
$(-\tfrac94 \calH + \calL ) + (\tfrac14 \calH ) = 
-2 \calH + \calL$.
The Hamiltonian matrix is 
\begin{equation} \label{eq:HA+1}
H_{F_z=+1}^{(A)} = 
\left(
\begin{array}{cc}
 \calL-2 \calH & \calV \\
 \calV & \calL-2 \calH \\
\end{array}
\right).
\end{equation}
The eigenvalues are
\begin{equation}
E^{(A)}_\pm = \calL - 2 \calH \pm \calV  \,,
\end{equation}
with the corresponding eigenvectors,
\begin{equation}
| u_\pm^{(A)} \rangle = \frac{1}{\sqrt{2}} \, 
\left( | \psi^{(A)}_{1} \rangle \pm | \psi^{(A)}_{2}\rangle \right) \,.
\end{equation}
Note that the designation of a degenerate subspace,
for the $F_z=+1$ subspace,
does not imply that there are no couplings to any other states
within the manifold; however, the couplings relating the 
degenerate states will become dominant for close approach.

A second degenerate subspace is given as
\begin{equation}
| \psi^{(B)}_{1} \rangle = | \psi^{\left(\mathrm{\Rmnum{2}}\right)}_{3} \rangle \,,
\qquad
| \psi^{(B)}_{2} \rangle = | \psi^{\left(\mathrm{\Rmnum{2}}\right)}_{5} \rangle \,.
\end{equation}
These states are composed of a triplet $S$ and a singlet
$P$ state, and hence the  diagonal entries in the
Hamiltonian matrix are
$(\tfrac34 \calH + \calL ) - (\tfrac34 \calH ) = \calL$.
The Hamiltonian matrix is
\begin{equation}
H_{F_z=+1}^{(B)} =
\left(
\begin{array}{cc}
 \calL & \calV \\
 \calV & \calL \\
\end{array}
\right).
\end{equation}
The eigenvalues are
\begin{equation}
E^{(B)}_\pm = \calL \pm \calV  \,,
\end{equation}
with the corresponding eigenvectors,
\begin{equation}
| u_\pm^{(B)} \rangle = \frac{1}{\sqrt{2}} \,
\left( | \psi^{(B)}_{1} \rangle \pm | \psi^{(B)}_{2} \right).
\end{equation}
The most complicated degenerate subspace is given by the 
vectors
\begin{align} \label{eq:PsiC+1}
| \psi^{(C)}_{1} \rangle =& \; | \psi^{\left(\mathrm{\Rmnum{2}}\right)}_{2} \rangle \,,
\qquad
| \psi^{(C)}_{2} \rangle = | \psi^{\left(\mathrm{\Rmnum{2}}\right)}_{4} \rangle \,,
\\[0.0077ex]
| \psi^{(C)}_{3} \rangle =& \; | \psi^{\left(\mathrm{\Rmnum{2}}\right)}_{6} \rangle \,,
\qquad
| \psi^{(C)}_{4} \rangle = | \psi^{\left(\mathrm{\Rmnum{2}}\right)}_{8} \rangle \,.
\end{align}
The Hamiltonian matrix is 
\begin{equation} \label{eq:HC+1}
H_{F_z=+1}^{(C)} =
\left(
\begin{array}{cccc}
 \calL+\calH & 0 & 0 & \calV \\
 0 & \calL+\calH & \calV & 0 \\
 0 & \calV & \calL+\calH & 0 \\
 \calV & 0 & 0 & \calL+\calH \\
\end{array}
\right) \,,
\end{equation}
which again decouples into two $2 \times 2$ matrices,
just like we saw in the case of $H_{F_z=+2}$.
The eigenvalues are 
\begin{equation}
E^{(C)}_\pm = \calH + \calL \pm \calV  \,,
\end{equation}
where the eigenvectors for $| u_{\pm,i}^{(C)} \rangle$ 
(with $i=1,2$ because of the degeneracy of the eigenvalues)
are given by 
\begin{subequations}
\begin{align}
| u_{\pm,1}^{(C)} \rangle &= \frac{1}{\sqrt{2}} \,
\left( | \psi^{(C)}_{1} \rangle \pm | \psi^{(C)}_{4} \right) \rangle\,,
\\
| u_{\pm,2}^{(C)} \rangle &= \frac{1}{\sqrt{2}} \,
\left( | \psi^{(C)}_{2} \rangle \pm | \psi^{(C)}_{3} \right) \rangle\,.
\end{align}
\end{subequations}

In Figs.~\ref{fig2} and~\ref{fig3}, we plot
the evolution of the eigenvalues of the matrices (\ref{matFz1Irr1}) and
(\ref{matFz1Irr2}) with respect to interatomic separation.
The larger energy shifts
within the subspace $\mathrm{\Rmnum{2}}$ are noticeable.
A feature
exhibited by the $F_z=+1$ manifold which was not present in the $F_z=+2$
manifold is that of level crossings: for sufficiently small interatomic
separation ($R<500\,a_0$), the eigenenergies of some of the states from the
submanifolds $\mathrm{\Rmnum{1}}$ and $\mathrm{\Rmnum{2}}$ 
in fact cross (these crossings would be visible if one were to 
superimpose Figs.~\ref{fig2} and~\ref{fig3}), 
while there are 
no level crossings between states belonging to the same submanifold.

\begin{figure*}[t!]
\begin{center}
\begin{minipage}{0.91\linewidth}
\begin{center}
\includegraphics[width=0.8\textwidth]{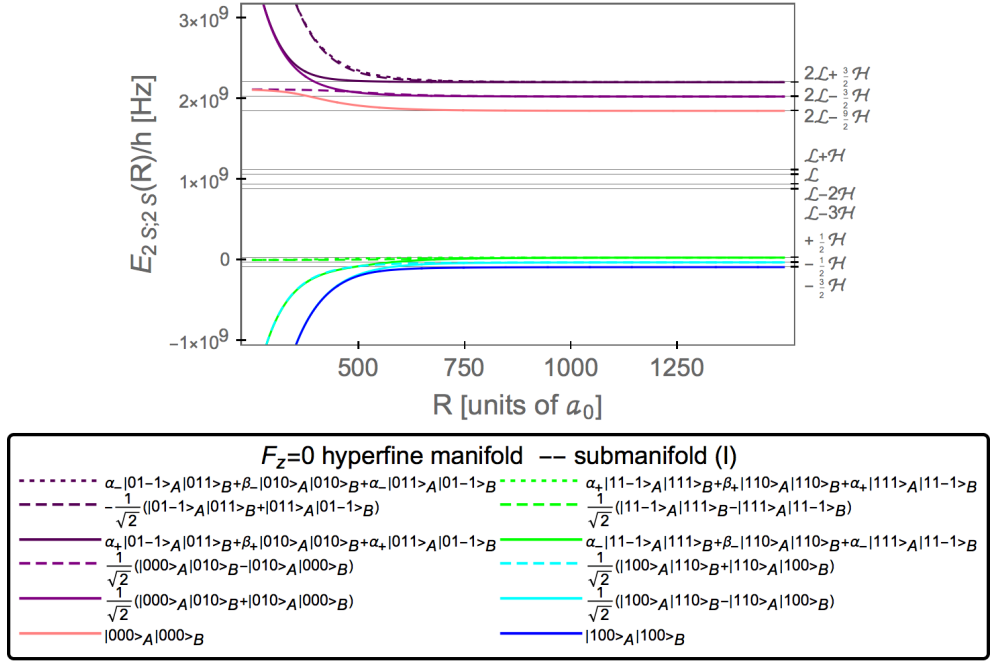}
\end{center}
\caption{(Color online.) Evolution of the energy levels of
the submanifold $\mathrm{\Rmnum{1}}$ within the $F_z=0$ hyperfine manifold as a
function of interatomic separation. 
Energetically, the $S$--$S$ states are above the 
$P$--$P$ states. The eigenstates given in the legend are
only asymptotic; for finite separation these states mix. Some of the curves
[namely, the third (from the top), sixth and twelfth] 
have been slightly offset for better readability. Notice
that, for sufficiently close separation ($R<1\,000\,a_0$), we witness some
level crossings between levels within the same submanifold $\mathrm{\Rmnum{1}}$.
The coefficients $\alpha_\pm$ and $\beta_\pm$ are determined by
second-order perturbation theory and given by Eq.~(\ref{eq:AlphaBeta}).
\label{fig4}}
\end{minipage}
\end{center}
\end{figure*}

\begin{figure*}[t!]
\begin{center}
\begin{minipage}{0.91\linewidth}
\begin{center}
\includegraphics[width=0.8\textwidth]{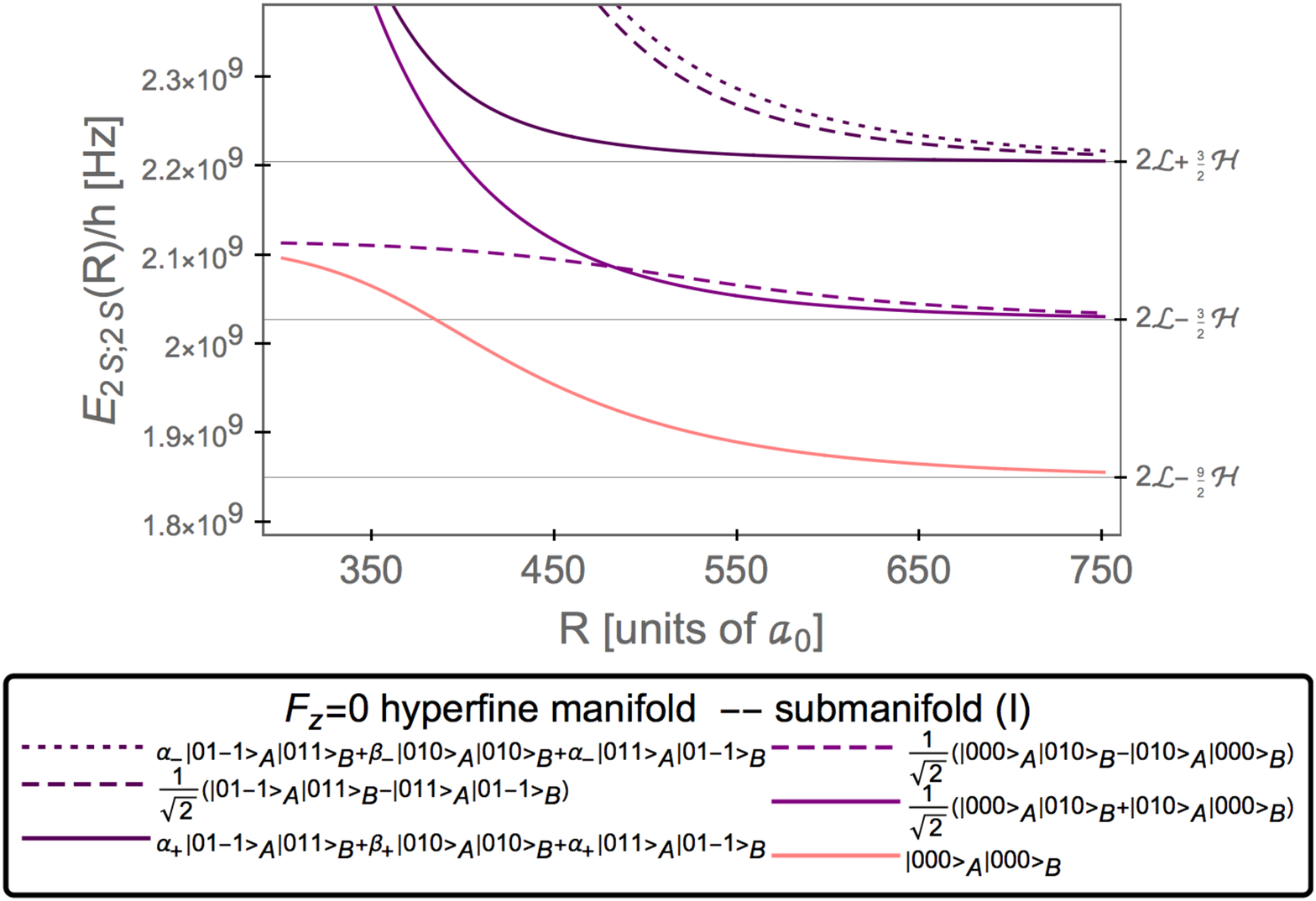}
\caption{(Color online.) Evolution of the energy levels of 
the $2S$--$2S$ states within the $F_z=0$ hyperfine manifold 
(subspace $\mathrm{\Rmnum{1}}$)
as a function of the interatomic separation 
(close-up of the ``upper'' levels in Fig.~\ref{fig4}).  The
eigenstates given in the legend are only asymptotic; for finite separation
these states mix. No offsets are used here. Notice that we witness one level
crossing. The coefficients $\alpha_\pm$ and $\beta_\pm$ are determined by
second-order perturbation theory and given by Eq.~(\ref{eq:AlphaBeta}).
\label{fig5}}
\end{center}
\end{minipage}
\end{center}
\end{figure*}

%
%
\subsection{Manifold $\maybebm{F_z = 0}$}
\label{sec33}

We can identify two irreducible subspaces within the $F_z=0$ manifold: the
subspace $\mathrm{\Rmnum{1}}$ is composed of 
states with both atoms in $S$, or both atoms in $P$ levels,
\begin{widetext}
\begin{equation}
\label{statesFz0Irr1}
\begin{aligned} [b]
| \Psi_{1}^{\left(\mathrm{\Rmnum{1}}\right)} \rangle =& \; 
| (0,0,0)_A \, (0,0,0)_B \rangle\,,  \qquad
| \Psi_{2}^{\left(\mathrm{\Rmnum{1}}\right)} \rangle = 
| (0,0,0)_A \, (0,1,0)_B \rangle\,, \qquad
| \Psi_{3}^{\left(\mathrm{\Rmnum{1}}\right)} \rangle = 
| (0,1,-1)_A \, (0,1,1)_B \rangle \,,
\\[0.0077ex]
| \Psi_{4}^{\left(\mathrm{\Rmnum{1}}\right)} \rangle =&\; 
| (0,1,0)_A \, (0,0,0)_B \rangle \,, \qquad
| \Psi_{5}^{\left(\mathrm{\Rmnum{1}}\right)} \rangle = 
| (0,1,0)_A \, (0,1,0)_B \rangle \,, \qquad
| \Psi_{6}^{\left(\mathrm{\Rmnum{1}}\right)} \rangle = 
| (0,1,1)_A \, (0,1,-1)_B \rangle \,,
\\[0.0077ex]
| \Psi_{7}^{\left(\mathrm{\Rmnum{1}}\right)} \rangle =& \;
| (1,0,0)_A \, (1,0,0)_B \rangle \,, \qquad
| \Psi_{8}^{\left(\mathrm{\Rmnum{1}}\right)} \rangle = 
| (1,0,0)_A \, (1,1,0)_B \rangle \,, \qquad
| \Psi_{9}^{\left(\mathrm{\Rmnum{1}}\right)} \rangle = 
| (1,1,-1)_A \, (1,1,1)_B \rangle \,,
\\[0.0077ex]
| \Psi_{10}^{\left(\mathrm{\Rmnum{1}}\right)} \rangle =& \;
| (1,1,0)_A \, (1,0,0)_B \rangle \,, \qquad
| \Psi_{11}^{\left(\mathrm{\Rmnum{1}}\right)} \rangle = 
| (1,1,0)_A \, (1,1,0)_B \rangle \,, \qquad
| \Psi_{12}^{\left(\mathrm{\Rmnum{1}}\right)} \rangle = 
| (1,1,1)_A \, (1,1,-1)_B \rangle \,
\end{aligned}
\end{equation}
and the Hamiltonian matrix reads
\begin{align}
\label{matFz0Irr1}
H_{F_z = 0}^{\left(\mathrm{\Rmnum{1}}\right)} =& \; \left(
\begin{array}{cccccccccccc}
 2 \calL-\tfrac92 \calH & 0 & 0 & 0 & 0 & 0 & 0 & 0 & - \calV & 0 & -2 \calV & - \calV \\
 0 & 2 \calL-\tfrac32 \calH & 0 & 0 & 0 & 0 & 0 & 0 & \calV & -2 \calV & 0 & - \calV \\
 0 & 0 & 2 \calL+\tfrac32 \calH & 0 & 0 & 0 & - \calV & \calV & 2 \calV & - \calV & \calV & 0 \\
 0 & 0 & 0 & 2 \calL-\tfrac32 \calH & 0 & 0 & 0 & -2 \calV & - \calV & 0 & 0 & \calV \\
 0 & 0 & 0 & 0 & 2 \calL+\tfrac32 \calH & 0 & -2 \calV & 0 & \calV & 0 & 0 & \calV \\
 0 & 0 & 0 & 0 & 0 & 2 \calL+\tfrac32 \calH & - \calV & - \calV & 0 & \calV & \calV & 2 \calV\\
 0 & 0 & - \calV & 0 & -2 \calV & - \calV & -\tfrac32 \calH & 0 & 0 & 0 & 0 & 0 \\
 0 & 0 & \calV & -2 \calV & 0 & - \calV & 0 & -\tfrac12 \calH & 0 & 0 & 0 & 0 \\
 - \calV & \calV & 2 \calV & - \calV & \calV & 0 & 0 & 0 & \tfrac12 \calH & 0 & 0 & 0 \\
 0 & -2 \calV & - \calV & 0 & 0 & \calV & 0 & 0 & 0 & -\tfrac12 \calH & 0 & 0 \\
 -2 \calV & 0 & \calV & 0 & 0 & \calV & 0 & 0 & 0 & 0 & \tfrac12 \calH & 0 \\
 - \calV & - \calV & 0 & \calV & \calV & 2 \calV & 0 & 0 & 0 & 0 & 0 & \tfrac12 \calH
\end{array}
\right) \,.
\end{align}
Subspace $\mathrm{\Rmnum{2}}$ is composed 
of the $S$--$P$ and $P$--$S$ combinations,
\begin{equation}
\label{statesFz0Irr2}
\begin{aligned} [b]
| \Psi_{1}^{\left(\mathrm{\Rmnum{2}}\right)} \rangle =& | (0,0,0)_A \, (1,0,0)_B \rangle\,, \qquad
| \Psi_{2}^{\left(\mathrm{\Rmnum{2}}\right)} \rangle = \; | (0,0,0)_A \, (1,1,0)_B \rangle \,, \qquad
| \Psi_{3}^{\left(\mathrm{\Rmnum{2}}\right)} \rangle = | (0,1,-1)_A \, (1,1,1)_B \rangle \,,
\\[0.0077ex]
| \Psi_{4}^{\left(\mathrm{\Rmnum{2}}\right)} \rangle =& | (0,1,0)_A \, (1,0,0)_B \rangle \,, \qquad
| \Psi_{5}^{\left(\mathrm{\Rmnum{2}}\right)} \rangle = \; | (0,1,0)_A \, (1,1,0)_B \rangle \,, \qquad
| \Psi_{6}^{\left(\mathrm{\Rmnum{2}}\right)} \rangle = | (0,1,1)_A \, (1,1,-1)_B \rangle \,,
\\[0.0077ex]
| \Psi_{7}^{\left(\mathrm{\Rmnum{2}}\right)} \rangle =& \; | (1,0,0)_A \, (0,0,0)_B \rangle \,, \qquad
| \Psi_{8}^{\left(\mathrm{\Rmnum{2}}\right)} \rangle = | (1,0,0)_A \, (0,1,0)_B \rangle \,, \qquad
| \Psi_{9}^{\left(\mathrm{\Rmnum{2}}\right)} \rangle = | (1,1,-1)_A \, (0,1,1)_B \rangle \,,
\\[0.0077ex]
| \Psi_{10}^{\left(\mathrm{\Rmnum{2}}\right)} \rangle =& \; | (1,1,0)_A \, (0,0,0)_B \rangle \,, \qquad
| \Psi_{11}^{\left(\mathrm{\Rmnum{2}}\right)} \rangle = | (1,1,0)_A \, (0,1,0)_B \rangle \,, \qquad
| \Psi_{12}^{\left(\mathrm{\Rmnum{2}}\right)} \rangle = | (1,1,1)_A \, (0,1,-1)_B \rangle \,,
\end{aligned}
\end{equation}
and the Hamiltonian matrix reads
\begin{align}
\label{matFz0Irr2}
H_{F_z = 0}^{\left(\mathrm{\Rmnum{2}}\right)} =& \; \left(
\begin{array}{cccccccccccc}
 \calL-3 \calH & 0 & 0 & 0 & 0 & 0 & 0 & 0 & - \calV & 0 & -2 \calV & - \calV \\
 0 & \calL-2 \calH & 0 & 0 & 0 & 0 & 0 & 0 & \calV & -2 \calV & 0 & - \calV \\
 0 & 0 & \calL+\calH & 0 & 0 & 0 & - \calV & \calV & 2 \calV & - \calV & \calV & 0 \\
 0 & 0 & 0 & \calL & 0 & 0 & 0 & -2 \calV & - \calV & 0 & 0 & \calV \\
 0 & 0 & 0 & 0 & \calL+\calH & 0 & -2 \calV & 0 & \calV & 0 & 0 & \calV \\
 0 & 0 & 0 & 0 & 0 & \calL+\calH & - \calV & - \calV & 0 & \calV & \calV & 2 \calV\\
 0 & 0 & - \calV & 0 & -2 \calV & - \calV & \calL-3 \calH & 0 & 0 & 0 & 0 & 0 \\
 0 & 0 & \calV & -2 \calV & 0 & - \calV & 0 & \calL & 0 & 0 & 0 & 0 \\
 - \calV & \calV & 2 \calV & - \calV & \calV & 0 & 0 & 0 & \calL+\calH & 0 & 0 & 0 \\
 0 & -2 \calV & - \calV & 0 & 0 & \calV & 0 & 0 & 0 & \calL-2 \calH & 0 & 0 \\
 -2 \calV & 0 & \calV & 0 & 0 & \calV & 0 & 0 & 0 & 0 & \calL+\calH & 0 \\
 - \calV & - \calV & 0 & \calV & \calV & 2 \calV & 0 & 0 & 0 & 0 & 0 & \calL+\calH
\end{array}
\right)\,.
\end{align}
\end{widetext}
Again, we notice that within the subspace $\mathrm{\Rmnum{1}}$, no two
degenerate levels are coupled to each other. On the other hand, the following
subspaces, within the subspace $\mathrm{\Rmnum{2}}$, can be identified as being
degenerate with respect to the unperturbed Hamiltonian, and having states
coupled by nonvanishing off-diagonal elements. 

\begin{figure*}[t!]
\begin{center}
\begin{minipage}{0.91\linewidth}
\begin{center}
\includegraphics[width=0.8\textwidth]{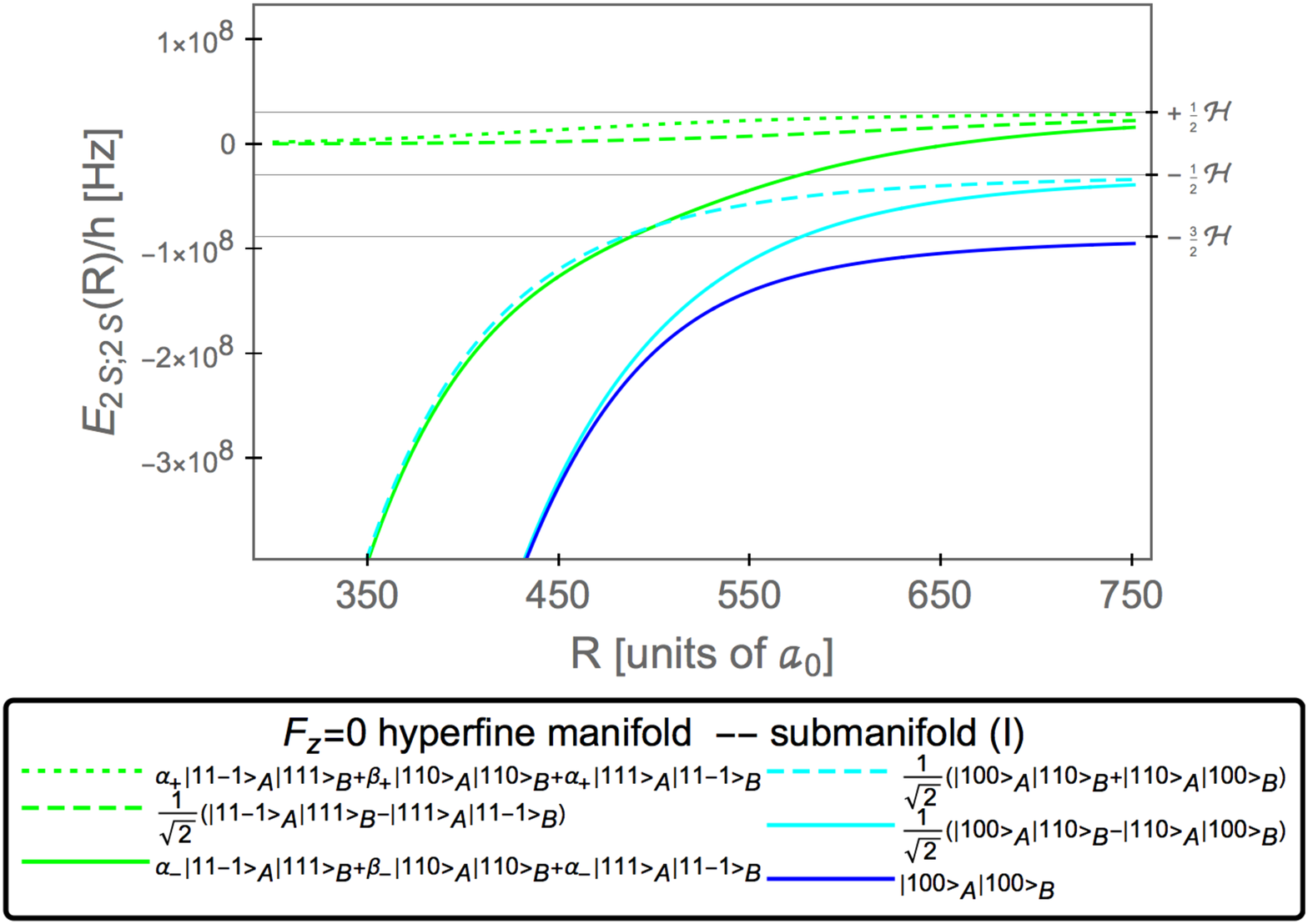}
\caption{(Color online.) Evolution of the energy levels of the $2P$--$2P$ 
states within the $F_z=0$ hyperfine manifold (subspace $\mathrm{\Rmnum{1}}$)
as a function of interatomic separation
(close-up of the ``lower'' levels in
Fig.~\ref{fig4}). Asymptotic 
eigenstates used in the legend mix for finite separation.
No offsets are used here. Notice that we witness one level
crossing. The coefficients $\alpha_\pm$ and $\beta_\pm$ are determined by
second-order perturbation theory and given by Eq.~(\ref{eq:AlphaBeta}).
\label{fig6}}
\end{center}
\end{minipage}
\end{center}
\end{figure*}

The first degenerate subspace
is given as follows,
\begin{equation}
| \Psi^{(A)}_{1} \rangle = | \Psi^{\left(\mathrm{\Rmnum{2}}\right)}_{2} \rangle \,,
\qquad
| \Psi^{(A)}_{2} \rangle = | \Psi^{\left(\mathrm{\Rmnum{2}}\right)}_{10} \rangle \,.
\end{equation}
The Hamiltonian matrix reads as 
\begin{equation}
H_{F_z=0}^{(A)} =
\left(
\begin{array}{cccc}
 \calL-2 \calH & -2 \calV \\
 -2 \calV & \calL-2 \calH \\
\end{array}
\right) \,.
\end{equation}
The eigensystem is given by
\begin{equation}
E^{(A)}_\pm = \calL - 2 \calH \pm 2 \calV\,,  \quad 
|u_\pm^{(A)}\rangle=\frac{1}{\sqrt{2}} \, 
( | \Psi_{1}^{(A)} \rangle \mp \Psi_{2}^{(A)} \rangle )\,.
\end{equation}
The second degenerate subspace is 
\begin{equation}
| \Psi^{(B)}_{1} \rangle = | \Psi^{\left(\mathrm{\Rmnum{2}}\right)}_{4} \rangle \,,
\qquad
| \Psi^{(B)}_{2} \rangle = | \Psi^{\left(\mathrm{\Rmnum{2}}\right)}_{8} \rangle \,,
\end{equation}
with the Hamiltonian matrix
\begin{equation}
H_{F_z=0}^{(B)} =
\left(
\begin{array}{cc}
 \calL & -2 \calV \\
 -2 \calV & \calL \\
\end{array}
\right)
\end{equation}
and the eigensystem
\begin{equation}
E^{(B)}_\pm = \calL \pm 2 \calV  \,, \quad \qquad 
|u_\pm^{(B)}\rangle=\frac{1}{\sqrt{2}} \, 
( | \Psi_{1}^{(B)} \rangle \mp \Psi_{2}^{(B)} \rangle )\,.
\end{equation}
The third degenerate subspace is
more complicated, and is spanned by the six state vectors
\begin{subequations}
\begin{align}
| \Psi^{(C)}_{1} \rangle =& \; | \Psi^{\left(\mathrm{\Rmnum{2}}\right)}_{3} \rangle \,,
\qquad
| \Psi^{(C)}_{2} \rangle = | \Psi^{\left(\mathrm{\Rmnum{2}}\right)}_{5} \rangle \,,
\\[0.0077ex]
| \Psi^{(C)}_{3} \rangle =& \; | \Psi^{\left(\mathrm{\Rmnum{2}}\right)}_{6} \rangle \,,
\qquad
| \Psi^{(C)}_{4} \rangle = | \Psi^{\left(\mathrm{\Rmnum{2}}\right)}_{9} \rangle \,,
\\[0.0077ex]
| \Psi^{(C)}_{5} \rangle =& \; | \Psi^{\left(\mathrm{\Rmnum{2}}\right)}_{11} \rangle \,,
\qquad
| \Psi^{(C)}_{6} \rangle = | \Psi^{\left(\mathrm{\Rmnum{2}}\right)}_{12} \rangle \,.
\end{align}
\end{subequations}
The six-dimensional submatrix is
\begin{widetext}
\begin{equation}
H_{F_z=0}^{(C)} =
\left(
\begin{array}{cccccc}
 \calL+\calH & 0 & 0 & 2 \calV & \calV & 0 \\
 0 & \calL+\calH & 0 & \calV & 0 & \calV \\
 0 & 0 & \calL+\calH & 0 & \calV & 2 \calV \\
 2 \calV & \calV & 0 & \calL+\calH & 0 & 0 \\
 \calV & 0 & \calV & 0 & \calL+\calH & 0 \\
 0 & \calV & 2 \calV & 0 & 0 & \calL+\calH \\
\end{array}
\right) \,.
\end{equation}
The eigenvalues are
\begin{subequations}
\begin{align}
E^{(C)}_{\pm,1} = & \; \calH + \calL \pm 2 \, \calV \,,
\\[0.0077ex]
E^{(C)}_{\pm,2} = & \; \calH + \calL \pm (\sqrt{3} + 1) \, \calV \,,
\\[0.0077ex]
E^{(C)}_{\pm,3} = & \; \calH + \calL \pm (\sqrt{3} - 1) \, \calV \,,
\end{align}
\end{subequations}
and the eigenvectors are
\begin{subequations} \label{eq:SixV}
\begin{align}
u^{(C)}_{+,1} = & \; \frac{1}{2} \,
\left( | \Psi^{(C)}_1 \rangle -
| \Psi^{(C)}_3 \rangle +
| \Psi^{(C)}_4 \rangle -
| \Psi^{(C)}_6 \rangle \right)  \,,
\\[0.0077ex]
u^{(C)}_{-,1} = & \; \frac{1}{2} \,
\left( | \Psi^{(C)}_1 \rangle -
| \Psi^{(C)}_3 \rangle -
| \Psi^{(C)}_4 \rangle +
| \Psi^{(C)}_6 \rangle \right)  \,,
\\[0.0077ex]
u^{(C)}_{+,2} = & \; \frac{1}{2 \sqrt{3-\sqrt{3}}} \,
\left( | \Psi^{(C)}_1 \rangle +
(\sqrt{3}-1) \, | \Psi^{(C)}_2 \rangle +
| \Psi^{(C)}_3 \rangle +
| \Psi^{(C)}_4 \rangle +
(\sqrt{3}-1) \, | \Psi^{(C)}_5 \rangle +
| \Psi^{(C)}_6 \rangle \right)  \,,
\\[0.0077ex]
u^{(C)}_{-,2} = & \; \frac{1}{2 \sqrt{3-\sqrt{3}}} \,
\left( | \Psi^{(C)}_1 \rangle +
(\sqrt{3}-1) \, | \Psi^{(C)}_2 \rangle +
| \Psi^{(C)}_3 \rangle -
| \Psi^{(C)}_4 \rangle -
(\sqrt{3}-1) \, | \Psi^{(C)}_5 \rangle -
| \Psi^{(C)}_6 \rangle \right)  \,,
\\[0.0077ex]
u^{(C)}_{+,3} = & \; \frac{1}{2 \sqrt{3+\sqrt{3}}} \,
\left( | \Psi^{(C)}_1 \rangle -
(\sqrt{3}+1) \, | \Psi^{(C)}_2 \rangle +
| \Psi^{(C)}_3 \rangle -
| \Psi^{(C)}_4 \rangle +
(\sqrt{3}+1) \, | \Psi^{(C)}_5 \rangle -
| \Psi^{(C)}_6 \rangle \right)  \,,
\\[0.0077ex]
u^{(C)}_{-,3} = & \; \frac{1}{2 \sqrt{3+\sqrt{3}}} \,
\left( | \Psi^{(C)}_1 \rangle -
(\sqrt{3}+1) \, | \Psi^{(C)}_2 \rangle +
| \Psi^{(C)}_3 \rangle +
| \Psi^{(C)}_4 \rangle -
(\sqrt{3}+1) \, | \Psi^{(C)}_5 \rangle +
| \Psi^{(C)}_6 \rangle \right)  \,.
\end{align}
\end{subequations}
\end{widetext}

\begin{figure*}[t!]
\begin{center}
\begin{minipage}{0.91\linewidth}
\begin{center}
\includegraphics[width=0.8\textwidth]{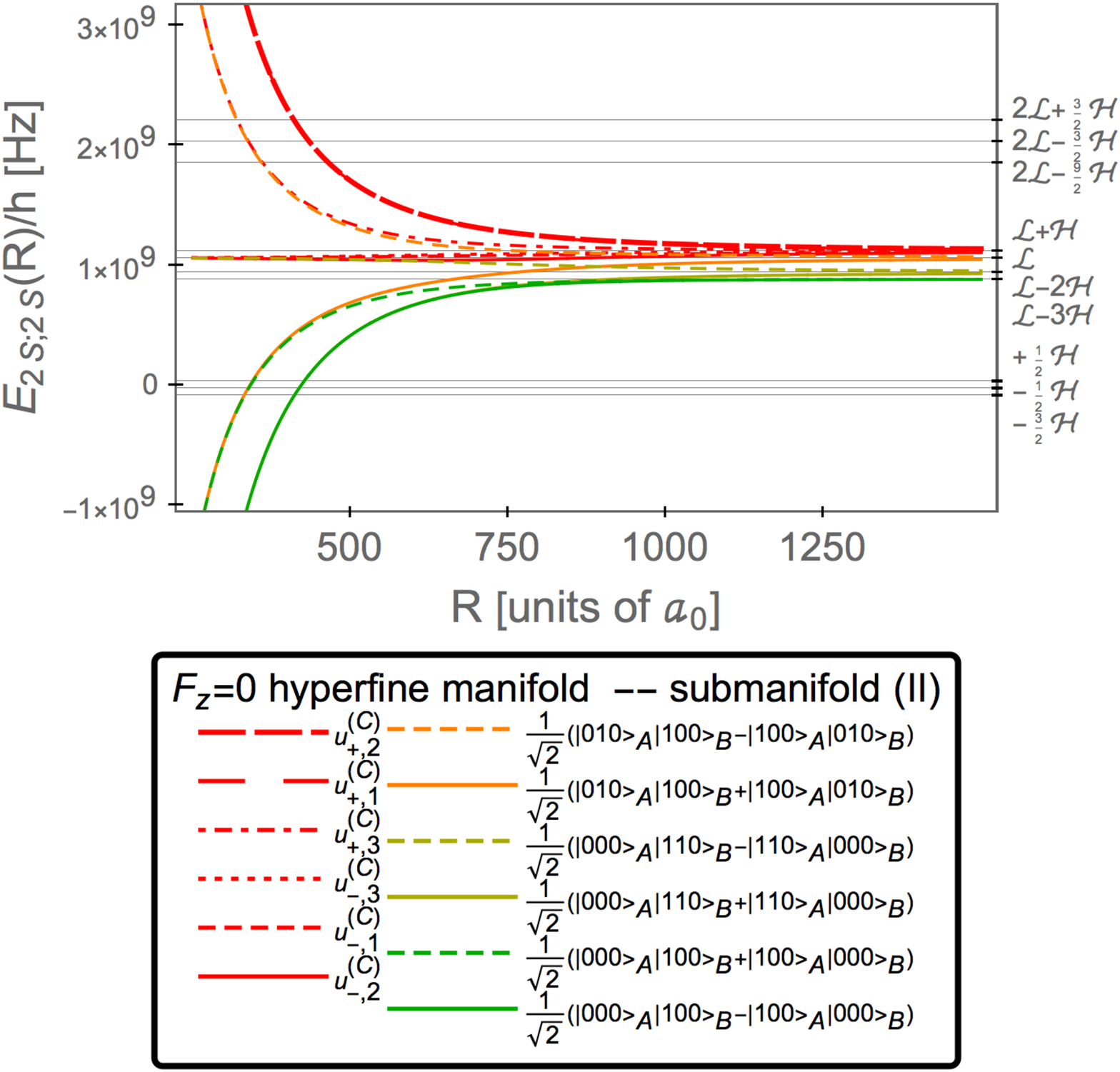}
\caption{(Color online.) Evolution of the $S$--$P$ and $P$--$S$ energy levels 
within the $F_z=0$ hyperfine manifold (submanifold $\mathrm{\Rmnum{2}}$)
as a function of interatomic separation.
The eigenstates given in the legend are only asymptotic, for finite separation
these states mix. Some of the curves (namely, for the ninth and
twelfth states in the legend, counted from the top) have been
slightly offset for better readability. Notice that, for sufficiently close
separation ($R<1\,000\,a_0$), we witness some level crossings between levels
within the same submanifold $\mathrm{\Rmnum{2}}$. The first six states 
are given in Eq.~(\ref{eq:SixV}). \label{fig7}}
\end{center}
\end{minipage}
\end{center}
\end{figure*}

\begin{figure*}[th]
\begin{center}
\begin{minipage}{0.91\linewidth}
\begin{center}
\includegraphics[width=0.8\textwidth]{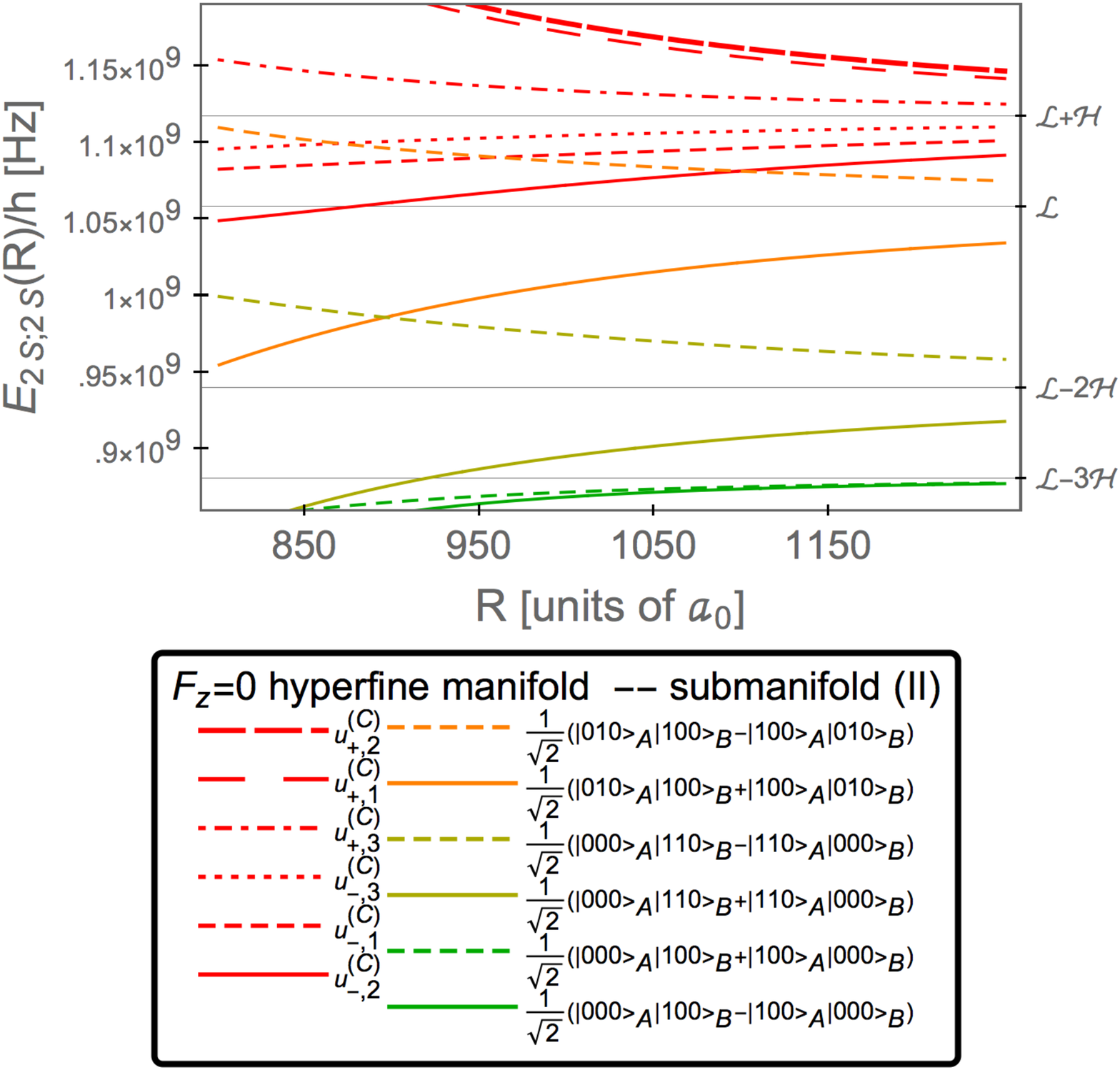}
\caption{(Color online.) 
Close-up of the evolution of the energy levels of the $2S$--$2P$ and $2P$--$2S$ states
(submanifold~$\mathrm{\Rmnum{2}}$) 
within the $F_z=0$ hyperfine manifold as a function of interatomic separation.
For the legend, we use the asymptotic eigenstates 
for large separation.
No offsets are used here. Notice that we witness four
level crossings. The first six states are given by
Eq.~(\ref{eq:SixV}).  \label{fig8}}
\end{center}
\end{minipage}
\end{center}
\end{figure*}

In Figs.~\ref{fig4}---\ref{fig8}, we plot the
evolution of the eigenvalues of matrices (\ref{matFz0Irr1}) and
(\ref{matFz0Irr2}) with respect to interatomic separation. Notice again that
the twelve levels within the subspace $\mathrm{\Rmnum{2}}$ noticeably leave
their asymptotic values (of order $\sim\calL$) for far larger separations
than the twelve levels within the subspace $\mathrm{\Rmnum{1}}$, as predicted
above by analyzing the order of the corresponding energy shifts. A feature
exhibited by the $F_z=0$ manifold which was not present in the $F_z=+1$
manifold is that of level crossings between levels within the same irreducible
submanifold: for sufficiently small interatomic separations ($R<1\,000\,a_0$),
the eigenenergies of some of the states from the submanifold
$\mathrm{\Rmnum{1}}$ cross between themselves, and so do some in manifold
$\mathrm{\Rmnum{2}}$. For better visibility of these intra-manifold crossings,
we present them in Figs.~\ref{fig5}
and~\ref{fig6}, as well as in Fig.~\ref{fig8}.  For even smaller
interatomic separations ($R<500\,a_0$) we obtain, again, crossings between
levels in manifolds $\mathrm{\Rmnum{1}}$ and $\mathrm{\Rmnum{2}}$.\\

As shown in Fig.~\ref{fig4}, some levels within 
submanifold $\mathrm{\Rmnum{1}}$, namely, on the one hand, the levels
\begin{subequations} 
\begin{align}
| \Psi_{3}^{\left(\mathrm{\Rmnum{1}}\right)} \rangle = & \;
| (0,1,-1)_A \, (0,1,1)_B \rangle \,, \\
| \Psi_{5}^{\left(\mathrm{\Rmnum{1}}\right)} \rangle = & \;
| (0,1,0)_A \, (0,1,0)_B \rangle \,, \\
| \Psi_{6}^{\left(\mathrm{\Rmnum{1}}\right)} \rangle = & \;
| (0,1,1)_A \, (0,1,-1)_B \rangle \,,
\end{align}
\end{subequations} 
that have asymptotic energy $2\calL+\tfrac{3}{2}\calH$; and, on the
other hand,
\begin{subequations} 
\begin{align}
| \Psi_{9}^{\left(\mathrm{\Rmnum{1}}\right)} \rangle = & \;
| (1,1,-1)_A \, (1,1,1)_B \rangle \,, \\
| \Psi_{11}^{\left(\mathrm{\Rmnum{1}}\right)} \rangle = & \;
| (1,1,0)_A \, (1,1,0)_B \rangle \,, \\
| \Psi_{12}^{\left(\mathrm{\Rmnum{1}}\right)} \rangle = & \;
| (1,1,1)_A \, (1,1,-1)_B \rangle \,,
\end{align}
\end{subequations} 
that have asymptotic energy $+\tfrac{1}{2}\calH$; are energetically
degenerate on the level of the unperturbed Hamiltonian, while experiencing no
first-order van der Waals couplings among themselves. They still split for
close enough interatomic distance because of higher-order couplings. This fixes
the coefficients $\alpha_\pm$ and $\beta_\pm$, according to the analysis
carried out in the following section
[see Fig.~\ref{fig4} and Eq.~(\ref{eq:AlphaBeta})].

%
%
\section{Hyperfine Shift in Specific Spectator States}
\label{sec4}

Of particular importance for hyperfine structure experiments
are energy differences of $2S$ singlet and triplet 
hyperfine sublevels, with the spectator atom in an arbitrary 
atomic state.
This amounts to the \vdw{} energy shift of the hyperfine lines,
i.e., the energy differences of the triplet level
$| (0,1,0)_A \, (\ell_B, F_B, F_{z,B})_B \rangle$
and the singlet level
$| (0,0,0)_A \, (\ell_B, F_B, F_{z,B})_B \rangle$,
for all possible states of atom $B$. We will see that the hyperfine 
frequencies are modified differently when the spectator 
atom is in a $2S$ or a $2P$ state.

Let us first examine the submanifold with $F_z=+1$.
The following states have the atom $A$ in the 
singlet hyperfine $2S$ level,
\begin{subequations}
\begin{align}
\label{stateFz1M1S1}
| \psi_{1}^{\left(\mathrm{\Rmnum{1}}\right)} \rangle &= 
| (0,0,0)_A \, (0,1,1)_B \rangle \,, \\
\label{stateFz1M2S1}
| \psi_{1}^{\left(\mathrm{\Rmnum{2}}\right)} \rangle &= 
| (0,0,0)_A \, (1,1,1)_B \rangle \,,
\end{align}
\end{subequations}
while 
\begin{subequations}
\begin{align}
\label{stateFz1M1S2}
| \psi_{2}^{\left(\mathrm{\Rmnum{1}}\right)} \rangle &= 
| (0,1,0)_A \, (0,1,1)_B \rangle \,, \\
\label{stateFz1M2S2}
| \psi_{2}^{\left(\mathrm{\Rmnum{2}}\right)} \rangle &= 
| (0,1,0)_A \, (1,1,1)_B \rangle \,, 
\end{align}
\end{subequations}
have the atom $A$ in the hyperfine triplet $S$ state.
The state of the spectator atom is preserved in the 
transitions $| \psi_{1}^{\left(\mathrm{\Rmnum{1}}\right)} \rangle \to 
| \psi_{2}^{\left(\mathrm{\Rmnum{1}}\right)} \rangle$
and $| \psi_{1}^{\left(\mathrm{\Rmnum{2}}\right)} \rangle \to 
| \psi_{2}^{\left(\mathrm{\Rmnum{2}}\right)} \rangle$.

For the states $| \psi_{1}^{\left(\mathrm{\Rmnum{2}}\right)} \rangle$ 
and $| \psi_{2}^{\left(\mathrm{\Rmnum{2}}\right)} \rangle$, 
the spectator atom is in a $P$ state. 
For both of these states, we can find energetically 
degenerate levels which are coupled to the reference state
by the \vdw{} interaction.
Specifically, $| \psi_{1}^{\left(\mathrm{\Rmnum{2}}\right)} \rangle$ is energetically 
degenerate with respect to 
$| \psi_{7}^{\left(\mathrm{\Rmnum{2}}\right)} \rangle = | (1,1,1)_A \, (0,0,0)_B \rangle$,
with the off-diagonal element
\begin{equation}
\langle \psi_{1}^{\left(\mathrm{\Rmnum{2}}\right)} | H_{\rm vdW} 
| \psi_{7}^{\left(\mathrm{\Rmnum{2}}\right)} \rangle = \calV \,.
\end{equation}
as can be seen in Eqs.~\eqref{IMPORTANT} and~\eqref{eq:HA+1}. Furthermore, 
$| \psi_{2}^{\left(\mathrm{\Rmnum{2}}\right)} \rangle$ is energetically 
degenerate with respect to 
$| \psi_{8}^{\left(\mathrm{\Rmnum{2}}\right)} \rangle = | (1,1,1)_A \, (0,1,0)_B \rangle$,
with the off-diagonal element
\begin{equation}
\langle \psi_{2}^{\left(\mathrm{\Rmnum{2}}\right)} | H_{\rm vdW} 
| \psi_{8}^{\left(\mathrm{\Rmnum{2}}\right)} \rangle = \calV
\end{equation}
as can be seen in Eqs.~\eqref{eq:PsiC+1} and~\eqref{eq:HC+1}. 
This implies that a hyperfine transition or energy difference,
with the spectator atom being in a $P$ state, 
undergoes a first-order \vdw{} energy shift proportional 
to $\calV$ [see Eq.~(\ref{eq:Inter})].

A close inspection of the matrix~\eqref{matFz1Irr1}
reveals that the levels 
$| \psi_{1}^{\left(\mathrm{\Rmnum{1}}\right)} \rangle$ and 
$| \psi_{2}^{\left(\mathrm{\Rmnum{1}}\right)} \rangle$
are not coupled to any energetically degenerate levels 
by the \vdw{} interaction;
hence, their leading-order shift is of second 
order in $\calV$.
From the previous analysis~\cite{AdEtAl2016vdWi} of the $(1S;nS)$ \vdw{} interaction,
however, we know that this observation does not 
imply that $| \psi_{1}^{\left(\mathrm{\Rmnum{1}}\right)} \rangle$ and 
$| \psi_{2}^{\left(\mathrm{\Rmnum{1}}\right)} \rangle$ decouple from 
any other levels in terms of the eigenstates of the 
total Hamiltonian $H$ given in Eq.~\eqref{H};
there may still be admixtures due to second-order 
effects in $H_{\rm vdW}$ which involve energetically 
degenerate levels, even if these are not coupled
directly to the reference state.
In the case of the $(1S;nS)$ \vdw{} interaction,
we had constructed an ``effective Hamiltonian''
$H_{\rm vdW} [1/(E_0 - H)]' H_{\rm vdW}$,
and evaluated its matrix elements in the 
basis of degenerate states.
The same approach is taken here, but with the Hamiltonian
matrix restricted to the relevant $F_z$ submanifold of states.

Let us illustrate the procedure.
We have the degenerate state
\begin{equation}
| \psi_{3}^{\left(\mathrm{\Rmnum{1}}\right)} \rangle = 
| (0,1,1)_A \, (0,0,0)_B \rangle \,, 
\end{equation}
which is obtained from $| \psi_{1}^{\left(\mathrm{\Rmnum{1}}\right)} \rangle$
by permuting the atoms $A$ and $B$,
and construct the restricted Hamiltonian matrix
\begin{equation}
\label{h13Def}
h_{1,3}^{\left(\mathrm{\Rmnum{1}}\right)} = \lim_{\epsilon\to 0} \left( \begin{array}{cc}
\langle \psi_{1}^{\left(\mathrm{\Rmnum{1}}\right)} | H^{(\epsilon)}_{\rm eff} | 
\psi_{1}^{\left(\mathrm{\Rmnum{1}}\right)} \rangle &
\langle \psi_{1}^{\left(\mathrm{\Rmnum{1}}\right)} | H^{(\epsilon)}_{\rm eff} | 
\psi_{3}^{\left(\mathrm{\Rmnum{1}}\right)} \rangle \\[1.0077ex]
\langle \psi_{3}^{\left(\mathrm{\Rmnum{1}}\right)} | H^{(\epsilon)}_{\rm eff} | 
\psi_{1}^{\left(\mathrm{\Rmnum{1}}\right)} \rangle &
\langle \psi_{3}^{\left(\mathrm{\Rmnum{1}}\right)} | H^{(\epsilon)}_{\rm eff} | 
\psi_{3}^{\left(\mathrm{\Rmnum{1}}\right)} \rangle 
\end{array} \right) \,.
\end{equation}
One defines the effective Hamiltonian $H_{\rm eff}^{\left(\epsilon\right)}$ as
follows. Let $H_1$ be the off-diagonal part of
$H_{F_z=+1}^{\left(\mathrm{\Rmnum{1}}\right)}$, equivalently given by the
expression of $H_{F_z=+1}^{\left(\mathrm{\Rmnum{1}}\right)}$ given in
Eq.~\eqref{matFz1Irr1} with $\calH \to 0$ and $\calL \to 0$.  Also,
let $H_0$ be the diagonal part of
$H_{F_z=+1}^{\left(\mathrm{\Rmnum{1}}\right)}$, equivalently given by the
expression of 
$H_{F_z=+1}^{\left(\mathrm{\Rmnum{1}}\right)}$ with $\calV \to 0$. Then
\begin{equation}
\label{Heff}
H_{\rm eff}^{\left(\epsilon\right)} = H_1 \cdot
\left( \frac{1}{E_{0,\psi_1^{\left(\mathrm{\Rmnum{1}}\right)}} - H_0 + \epsilon} 
\right) \cdot H_1  \,,
\end{equation}
where the dot (``$\cdot$'') denotes the matrix multiplication
and the Green function matrix 
$[1/(E_{0,\psi_1^{\left(\mathrm{\Rmnum{1}}\right)}} - H_0 + \epsilon)]$
is obtained as the inverse of the diagonal 
matrix $\mathbbm{1} E_{0,\psi_1^{\left(\mathrm{\Rmnum{1}}\right)}} - H_0 =
\mathbbm{1} E_{0,\psi_3^{\left(\mathrm{\Rmnum{1}}\right)}} - H_0$.
Since $\langle \psi_{1}^{\left(\mathrm{\Rmnum{1}}\right)} | H_1 
| \psi_{3}^{\left(\mathrm{\Rmnum{1}}\right)} \rangle = 0$,
it is not necessary to use the reduced Green function (which excludes degenerate states);
the limit $\epsilon \to 0$ is finite for all elements in 
$h_{1,3}^{\left(\mathrm{\Rmnum{1}}\right)}$.
The matrix $h_{1,3}^{\left(\mathrm{\Rmnum{1}}\right)}$ takes the following form,
\begin{equation} \label{eq:h13}
h_{1,3}^{\left(\mathrm{\Rmnum{1}}\right)} = \left( \begin{array}{cc}
\dfrac52 \, \dfrac{\calV^2}{\calL - \calH} + 
\dfrac{\calV^2}{2 \calL - \calH} &
\dfrac{2 \calV^2}{\calL - \calH} \\[2ex]
\dfrac{2 \calV^2}{\calL - \calH} & 
\dfrac52 \, \dfrac{\calV^2}{\calL - \calH} + 
\dfrac{\calV^2}{2 \calL - \calH} 
\end{array} \right) \,,
\end{equation}
with eigenvalues
\begin{equation}
\label{eps13}
\epsilon_{1,3}^{\left(\mathrm{\Rmnum{1}}\right)\pm} = 
\dfrac52 \, \dfrac{\calV^2}{\calL - \calH} + 
\dfrac{\calV^2}{2 \calL - \calH}
\pm \dfrac{2 \calV^2}{\calL - \calH} \,,
\end{equation}
akin to the formula $C_6 = D_6 \pm M_6$ encountered in Ref.~\cite{AdEtAl2016vdWi},
with eigenvectors 
\begin{equation}
\label{mark1}
|\psi_{1,3}^{\left(\mathrm{\Rmnum{1}}\right)\pm} \rangle = 
\frac{1}{\sqrt{2}} \, 
\left( | \psi_1^{\left(\mathrm{\Rmnum{1}}\right)} \rangle \pm 
| \psi_3^{\left(\mathrm{\Rmnum{1}}\right)} \rangle \right) \,.
\end{equation}
Note that the eigenvalues $\epsilon_{1,3}^\pm$ only refer to the 
interaction energy; in order to obtain the 
eigenvalue of the total Hamiltonian $H$ given in Eq.~\eqref{H},
one has to add the unperturbed entry 
$2 \calL-\tfrac32 \calH$.

For the reference state $| \psi_2^{\left(\mathrm{\Rmnum{1}}\right)}\rangle$, 
we have the degenerate
state $| \psi_{4}^{\left(\mathrm{\Rmnum{1}}\right)} \rangle = 
| (0,1,1)_A \, (0,1,0)_B \rangle$ [see Eq.~(\ref{statesFz1Irr1})].
The matrix $h_{2,4}^{\left(\mathrm{\Rmnum{1}}\right)}$ 
has the same structure as (but different elements from) 
$h_{1,3}^{\left(\mathrm{\Rmnum{1}}\right)}$
given in Eq.~\eqref{eq:h13}, and we find [see Eq.~(\ref{matFz1Irr1})]

\begin{equation}
\label{eps24}
\epsilon_{2,4}^{\left(\mathrm{\Rmnum{1}}\right)\pm} = 
\dfrac52 \, \dfrac{\calV^2}{\calL + \calH} + \dfrac{\calV^2}{2 \calL + \calH}
\pm \dfrac{2 \calV^2}{\calL + \calH} \,.
\end{equation}
The expression for $\epsilon_{2,4}^\pm$ is obtained from 
$\epsilon_{1,3}^\pm$ by a sign change in $\calH$.
The eigenvectors are
\begin{equation}
\label{mark2}
|\psi_{2,4}^{\left(\mathrm{\Rmnum{1}}±\right)\pm} \rangle =
\frac{1}{\sqrt{2}} \,
\left( | \psi_2^{\left(\mathrm{\Rmnum{1}}\right)} \rangle 
\pm | \psi_4^{\left(\mathrm{\Rmnum{1}}\right)} \rangle \right) \,.
\end{equation}
The unperturbed energy for the states
$| \psi_{2}^{\left(\mathrm{\Rmnum{1}}\right)} \rangle$ and 
$| \psi_{4}^{\left(\mathrm{\Rmnum{1}}\right)} \rangle$ is
$2 \calL+\tfrac32 \calH$.
Hence, in the transition
$| \psi_{1}^{\left(\mathrm{\Rmnum{1}}\right)} \rangle \to
| \psi_{2}^{\left(\mathrm{\Rmnum{1}}\right)} \rangle$,
where both atoms are in $S$ states,
one has only second-order \vdw{} shifts.
We recall that the transition is 
$| (0,0,0)_A \, (0,1,1)_B \rangle \to 
| (0,1,0)_A \, (0,1,1)_B \rangle$.

We also need to analyze the space with $F_z = 0$.
The following states have the atom $A$ in the
singlet hyperfine $2S$ level,
\begin{subequations}
\begin{align}
\label{stateFz0M1S1}
| \Psi_{1}^{\left(\mathrm{\Rmnum{1}}\right)} \rangle &= 
| (0,0,0)_A \, (0,0,0)_B \rangle\,,  \\[0.0077ex]
\label{stateFz0M1S2}
| \Psi_{2}^{\left(\mathrm{\Rmnum{1}}\right)} \rangle &= 
| (0,0,0)_A \, (0,1,0)_B \rangle\,, \\[0.0077ex]
\label{stateFz0M2S1}
| \Psi_{1}^{\left(\mathrm{\Rmnum{2}}\right)} \rangle &= 
| (0,0,0)_A \, (1,0,0)_B \rangle\,, \\[0.0077ex]
\label{stateFz0M2S2}
| \Psi_{2}^{\left(\mathrm{\Rmnum{2}}\right)} \rangle &= 
| (0,0,0)_A \, (1,1,0)_B \rangle \,,
\end{align}
\end{subequations}
while the $2S$ hyperfine triplet state of atom $A$ is present in 
the states
\begin{subequations}
\begin{align}
\label{stateFz0M1S4}
| \Psi_{4}^{\left(\mathrm{\Rmnum{1}}\right)} \rangle &= 
| (0,1,0)_A \, (0,0,0)_B \rangle \,, \\[0.0077ex]
\label{stateFz0M1S5}
| \Psi_{5}^{\left(\mathrm{\Rmnum{1}}\right)} \rangle &= 
| (0,1,0)_A \, (0,1,0)_B \rangle \,, \\[0.0077ex]
\label{stateFz0M2S4}
| \Psi_{4}^{\left(\mathrm{\Rmnum{2}}\right)} \rangle &= 
| (0,1,0)_A \, (1,0,0)_B \rangle \,, \\[0.0077ex]
\label{stateFz0M2S5}
| \Psi_{5}^{\left(\mathrm{\Rmnum{2}}\right)} \rangle &= 
| (0,1,0)_A \, (1,1,0)_B \rangle \,.
\end{align}
\end{subequations}
The transitions in question are
$| \Psi_{1}^{\left(\mathrm{\Rmnum{1}}\right)} \rangle \to 
| \Psi_{4}^{\left(\mathrm{\Rmnum{1}}\right)} \rangle$,
$| \Psi_{2}^{\left(\mathrm{\Rmnum{1}}\right)} \rangle \to 
| \Psi_{5}^{\left(\mathrm{\Rmnum{1}}\right)} \rangle$,
$| \Psi_{1}^{\left(\mathrm{\Rmnum{2}}\right)} \rangle \to 
| \Psi_{4}^{\left(\mathrm{\Rmnum{2}}\right)} \rangle$,
and
$| \Psi_{2}^{\left(\mathrm{\Rmnum{2}}\right)} \rangle \to 
| \Psi_{5}^{\left(\mathrm{\Rmnum{2}}\right)} \rangle$.
In view of the results
\begin{equation}
\langle \Psi_{4}^{\left(\mathrm{\Rmnum{2}}\right)} | 
H_{\rm vdW} | \Psi_{8}^{\left(\mathrm{\Rmnum{2}}\right)} \rangle =
\langle \Psi_{2}^{\left(\mathrm{\Rmnum{2}}\right)} | 
H_{\rm vdW} | \Psi_{10}^{\left(\mathrm{\Rmnum{2}}\right)} \rangle = -2 \calV 
\end{equation}
and
\begin{equation}
\langle \Psi_{5}^{\left(\mathrm{\Rmnum{2}}\right)} | 
H_{\rm vdW} | \Psi_{9}^{\left(\mathrm{\Rmnum{2}}\right)} \rangle =
\langle \Psi_{5}^{\left(\mathrm{\Rmnum{2}}\right)} | 
H_{\rm vdW} | \Psi_{12}^{\left(\mathrm{\Rmnum{2}}\right)} \rangle = -2 \calV \,,
\end{equation}
which we obtain from Eq.~(\ref{matFz0Irr2}), both transitions 
$| \Psi_{1}^{\left(\mathrm{\Rmnum{2}}\right)} \rangle \to 
| \Psi_{4}^{\left(\mathrm{\Rmnum{2}}\right)} \rangle$,
and $| \Psi_{2}^{\left(\mathrm{\Rmnum{2}}\right)} \rangle \to 
| \Psi_{5}^{\left(\mathrm{\Rmnum{2}}\right)} \rangle$
undergo first-order \vdw{} shifts. The spectator atom in these
cases is in a $P$ state.

By contrast, for the transitions within the submanifold $\mathrm{\Rmnum{1}}$,
namely, $| \Psi_{1}^{\left(\mathrm{\Rmnum{1}}\right)} \rangle \to |
\Psi_{4}^{\left(\mathrm{\Rmnum{1}}\right)} \rangle$ and 
$| \Psi_{2}^{\left(\mathrm{\Rmnum{1}}\right)} \rangle \to 
| \Psi_{5}^{\left(\mathrm{\Rmnum{1}}\right)} \rangle$, the \vdw{} shift only
enters in second order.  We first analyze the transition 
$| \Psi_{1}^{\left(\mathrm{\Rmnum{1}}\right)} \rangle \to |
\Psi_{4}^{\left(\mathrm{\Rmnum{1}}\right)} \rangle = | (0,0,0)_A \, (0,0,0)_B
\rangle \to | (0,1,0)_A \, (0,0,0)_B \rangle$.  There is no energetically
degenerate state available for 
$| \Psi_{1}^{\left(\mathrm{\Rmnum{1}}\right)} \rangle $, and hence
one obtains
\begin{equation}
\label{EFz0M1S1}
\Delta E_{\Psi_1^{\left(\mathrm{\Rmnum{1}}\right)}} = 
\frac{6 \calV^2}{{\color{black}2}\calL - 5 \calH} 
{\color{black}+\left(
\frac{\calV^2}{2 \calL - \calH} \right)}
\end{equation}
from Eq.~(\ref{matFz0Irr1}). The levels 
$| \Psi_{2}^{\left(\mathrm{\Rmnum{1}}\right)} \rangle$ and 
$| \Psi_{4}^{\left(\mathrm{\Rmnum{1}}\right)} \rangle$
are energetically degenerate with respect to their 
unperturbed energy $2 \calL-\tfrac32 \calH$,
but there is no direct 
\vdw{} coupling between them.
The matrix $H_{2,4}^{\mathrm{\Rmnum{1}}}$ is easily calculated in analogy 
to $h_{1,3}^{\mathrm{\Rmnum{1}}}$ given in Eq.~\eqref{eq:h13}, the difference being 
that the effective interaction Hamiltonian~\eqref{Heff} needs to be calculated 
with respect to $H_{F_z=0}$, not $H_{F_z=+1}$.
We find the eigenvalues
\begin{equation}
\label{E24}
E_{2,4}^{\left(\mathrm{\Rmnum{1}}\right)\pm} = \dfrac{\calV^2}{\calL - \calH} 
+ \dfrac{4 \calV^2}{2 \calL - \calH}
\pm \dfrac{\calV^2}{-\calL + \calH} \,.
\end{equation}
with eigenvectors
\begin{equation}
|\Psi_{2,4}^{\left(\mathrm{\Rmnum{1}}\right)\pm} \rangle =
\frac{1}{\sqrt{2}} \,
\left( | \Psi_2^{\left(\mathrm{\Rmnum{1}}\right)} \rangle \pm 
| \Psi_4^{\left(\mathrm{\Rmnum{1}}\right)} \rangle \right) \,.
\end{equation}
The last state whose \vdw{} interaction energy needs to 
be analyzed is $| \Psi_5^{\left(\mathrm{\Rmnum{1}}\right)} \rangle$.
This state forms a degenerate set together with the states 
$| \Psi_{3}^{\left(\mathrm{\Rmnum{1}}\right)} \rangle$ 
and $| \Psi_{6}^{\left(\mathrm{\Rmnum{1}}\right)} \rangle$,
\begin{subequations}
\begin{align}
| \Psi_{3}^{\left(\mathrm{\Rmnum{1}}\right)} \rangle =& \; 
| (0,1,-1)_A \, (0,1,1)_B \rangle \,, 
\\[2ex]
| \Psi_{5}^{\left(\mathrm{\Rmnum{1}}\right)} \rangle =& \; 
| (0,1,0)_A \, (0,1,0)_B \rangle \,, 
\\[2ex]
| \Psi_{6}^{\left(\mathrm{\Rmnum{1}}\right)} \rangle =& \; 
| (0,1,1)_A \, (0,1,-1)_B \rangle \,,
\end{align}
\end{subequations}
which are both composed of two hyperfine triplet $S$ states.
Under the additional approximation $\calH \ll \calL$,
one finds through Eq.~(\ref{matFz0Irr1}) the Hamiltonian matrix
\begin{equation}
\label{H356}
H_{3,5,6}^{\left(\mathrm{\Rmnum{1}}\right)} \approx \left( \begin{array}{ccc}
\dfrac{4 \calV^2}{\calL} & \dfrac{2 \calV^2}{\calL} & 0 \\[2ex]
\dfrac{2 \calV^2}{\calL} & \dfrac{3 \calV^2}{\calL} & \dfrac{2 \calV^2}{\calL} 
\\[2ex]
0 & \dfrac{2 \calV^2}{\calL} & \dfrac{4 \calV^2}{\calL}  
\end{array} \right) \,.
\end{equation}
The energy eigenvalues are
\begin{subequations}
\label{E356}
\begin{align}
E^{\left(\mathrm{\Rmnum{1}}\right)\left(1\right)}_{3,5,6} \approx & \; 
\frac{7 + \sqrt{33}}{2} \, \frac{\calV^2}{\calL} \,,
\\[0.0077ex]
E^{\left(\mathrm{\Rmnum{1}}\right)\left(2\right)}_{3,5,6} \approx & \; 
\frac{4 \calV^2}{\calL} \,,
\\[0.0077ex]
E^{\left(\mathrm{\Rmnum{1}}\right)\left(3\right)}_{3,5,6} \approx & \; 
\frac{7 - \sqrt{33}}{2} \, \frac{\calV^2}{\calL} \,,
\end{align}
\end{subequations}
with eigenvectors
\begin{subequations}
\begin{align}
\Psi^{\left(\mathrm{\Rmnum{1}}\right)\left(1\right)}_{3,5,6} \approx & \; 
\alpha_- \, |\Psi_3^{\left(\mathrm{\Rmnum{1}}\right)}\rangle +
\beta_- \, |\Psi_5^{\left(\mathrm{\Rmnum{1}}\right)}\rangle+\alpha_- \, 
|\Psi_6^{\left(\mathrm{\Rmnum{1}}\right)} \rangle \,,
\\[0.0077ex]
\Psi^{\left(\mathrm{\Rmnum{1}}\right)\left(2\right)}_{3,5,6} \approx & \; 
- \frac{1}{\sqrt{2}} |\Psi_3^{\left(\mathrm{\Rmnum{1}}\right)}\rangle +
 \frac{1}{\sqrt{2}} |\Psi_6^{\left(\mathrm{\Rmnum{1}}\right)}\rangle \,,
\\[0.0077ex]
\Psi^{\left(\mathrm{\Rmnum{1}}\right)\left(3\right)}_{3,5,6} \approx & \; 
\alpha_+ \, |\Psi_3^{\left(\mathrm{\Rmnum{1}}\right)}\rangle +\beta_+ \, 
|\Psi_5^{\left(\mathrm{\Rmnum{1}}\right)}\rangle +
\alpha_+ \, |\Psi_6^{\left(\mathrm{\Rmnum{1}}\right)} \rangle \,
\end{align}
\end{subequations}
where we introduced the notation
\begin{subequations} 
\label{eq:AlphaBeta}
\begin{align}
\alpha_\pm&=2\sqrt{\frac{2}{33\pm\sqrt{33}}}\,,\\
\beta_\pm&=\mp\frac{\sqrt{33}\pm1}{\sqrt{2\left(33\pm\sqrt{33}\right)}}\,.
\end{align}
\end{subequations} 
The transitions 
$| \Psi_{1}^{\left(\mathrm{\Rmnum{1}}\right)} \rangle \to 
| \Psi_{4}^{\left(\mathrm{\Rmnum{1}}\right)} \rangle$
and
$| \Psi_{2}^{\left(\mathrm{\Rmnum{1}}\right)} \rangle \to 
| \Psi_{5}^{\left(\mathrm{\Rmnum{1}}\right)} \rangle$
thus undergo only second-order \vdw{} shifts
of order $\calV^2/\calL$;
these are the only hyperfine transitions with both 
atoms in metastable states.

Finally, we briefly mention the $F_z=-1$ subspace. The analysis carried out for
the $F_z=+1$ subspace holds when we perform the substitutions [see also
Appendix~\ref{appa1}]
\begin{equation} \label{eq:SubstiSubOne}
\begin{aligned} [b]
| \psi_{1}^{\left(\mathrm{\Rmnum{1}}\right)} \rangle &\rightarrow| 
\psi_{1}^{\prime\left(\mathrm{\Rmnum{1}}\right)} \rangle\,,\quad
| \psi_{2}^{\left(\mathrm{\Rmnum{1}}\right)} \rangle \rightarrow| 
\psi_{4}^{\prime\left(\mathrm{\Rmnum{1}}\right)} \rangle\,,\quad
| \psi_{3}^{\left(\mathrm{\Rmnum{1}}\right)} \rangle \rightarrow| 
\psi_{2}^{\prime\left(\mathrm{\Rmnum{1}}\right)} \rangle\,,\\
| \psi_{4}^{\left(\mathrm{\Rmnum{1}}\right)} \rangle &\rightarrow| 
\psi_{3}^{\prime\left(\mathrm{\Rmnum{1}}\right)} \rangle\,,\quad
| \psi_{5}^{\left(\mathrm{\Rmnum{1}}\right)} \rangle \rightarrow| 
\psi_{5}^{\prime\left(\mathrm{\Rmnum{1}}\right)} \rangle\,,\quad
| \psi_{6}^{\left(\mathrm{\Rmnum{1}}\right)} \rangle \rightarrow| 
\psi_{8}^{\prime\left(\mathrm{\Rmnum{1}}\right)} \rangle\,,\\
| \psi_{7}^{\left(\mathrm{\Rmnum{1}}\right)} \rangle &\rightarrow| 
\psi_{6}^{\prime\left(\mathrm{\Rmnum{1}}\right)} \rangle\,,\quad
| \psi_{8}^{\left(\mathrm{\Rmnum{1}}\right)} \rangle \rightarrow| 
\psi_{7}^{\prime\left(\mathrm{\Rmnum{1}}\right)} \rangle\,.
\end{aligned}
\end{equation}

In Tables~\ref{tab:Epsilon1},~\ref{tab:Epsilon2},~\ref{tab:E1}
and~\ref{tab:E2}, we provide some numerical values for the modification of the
$2S$ hyperfine splitting, as a function of interatomic distance. The spectator
atom is in an $S$ state for Tables~\ref{tab:Epsilon1} and~\ref{tab:E1} and in a
$P$ state for Tables~\ref{tab:Epsilon2} and~\ref{tab:E2}.
Tables~\ref{tab:Epsilon1} and \ref{tab:Epsilon2} treat of the relevant
transitions within the $F_z=+1$ manifold, while Tables~\ref{tab:E1}
and~\ref{tab:E2} treat of (some of) the relevant transitions within the $F_z=0$
manifold. The relevant transitions within the $F_z=-1$ manifold have the same
transition energies as those within the $F_z=+1$ for all separations, and the
corresponding results can thusly be read from Tables~\ref{tab:Epsilon1}
and~\ref{tab:Epsilon2}, with the substitutions
\begin{subequations} \label{eq:TableSubst}
\begin{align}
| (0,0,0)_A \, (0,1,1)_B \rangle&\rightarrow| (0,0,0)_A \, (0,1,-1)_B \rangle\,,\\
| (0,1,0)_A \, (0,1,1)_B \rangle&\rightarrow| (0,1,0)_A \, (0,1,-1)_B \rangle\,,\\
| (0,0,0)_A \, (1,1,1)_B \rangle&\rightarrow| (0,0,0)_A \, (1,1,-1)_B \rangle\,,\\
| (0,1,0)_A \, (1,1,1)_B \rangle&\rightarrow| (0,1,0)_A \, (1,1,-1)_B \rangle\,.
\end{align}
\end{subequations}

\begin{center}
\def\arraystretch{1.25}
\begin{table} [t]
\begin{center}
\begin{tabular}{c@{\hspace{0.3cm}}S[table-format=-1.4e-1]S[table-format=-1.4e-1]}
\hline
\hline
R & 
\multicolumn{1}{c}{$\Delta\epsilon_+^{\left(\mathrm{\Rmnum{1}}\right)}$} & 
\multicolumn{1}{c}{$\Delta\epsilon_-^{\left(\mathrm{\Rmnum{1}}\right)}$} \\
\hline
$\infty$&    0         &  0         \\
$750\,a_0$& -1.1099e-2 & -2.1156e-3 \\
$500\,a_0$&  5.5547e-1 &  6.8788e-2 \\
$250\,a_0$&  3.7979e1  &  2.5507e1  \\
\hline
\hline
\end{tabular}
\end{center}
\caption{Energy shifts with the spectator atom in an $S$ state with $F=1$:
numerical values of the \vdw{} shift to the energy difference
$\Delta\epsilon_+^{\left(\mathrm{\Rmnum{1}}\right)}$ between the symmetric
superpositions 
$\left(1/\sqrt{2}\right)\left(| (0,1,0)_A \, (0,1,1)_B \rangle+
| (0,1,1)_A \, (0,1,0)_B \rangle\right)$ and 
$\left(1/\sqrt{2}\right)\left(| (0,0,0)_A \, (0,1,1)_B \rangle+
| (0,1,1)_A \, (0,0,0)_B \rangle\right)$, and to
the energy difference 
$\Delta\epsilon_-^{\left(\mathrm{\Rmnum{1}}\right)}$
between the antisymmetric superpositions 
$\left(1/\sqrt{2}\right)\left(| (0,1,0)_A \, (0,1,1)_B \rangle-
| (0,1,1)_A \, (0,1,0)_B \rangle\right)$ and
$\left(1/\sqrt{2}\right)\left(| (0,0,0)_A \, (0,1,1)_B \rangle-
| (0,1,1)_A \, (0,0,0)_B \rangle\right)$; 
as a function of the interatomic separation $R$. 
We recall that the asymptotic value of these energy differences is given by $3\calH$;
the unperturbed energies are $2 \calL \pm \tfrac32 \calH$
[see the text surrounding Eqs.~\eqref{mark1} and~\eqref{mark2}]. All
energies are given in units of the hyperfine splitting constant $\calH$
defined by (\ref{eq:HFSplit}). \label{tab:Epsilon1}}
\end{table}
\end{center}
 
\begin{center}
\def\arraystretch{1.25}
\begin{table} [t]
\begin{center}
\begin{tabular}{c@{\hspace{0.3cm}}S[table-format=-1.4e-1]S[table-format=-1.4e-1]}
\hline
\hline
R & 
\multicolumn{1}{c}{$\Delta\epsilon_+^{\left(\mathrm{\Rmnum{2}}\right)}$} &
\multicolumn{1}{c}{$\Delta\epsilon_-^{\left(\mathrm{\Rmnum{2}}\right)}$} \\
\hline
$\infty$   & 0        & 0        \\
$750\,a_0$ & 2.6396   & 2.6396   \\
$500\,a_0$ & 1.3276e1 & 1.3276e1 \\
$250\,a_0$ & 1.2510e2 & 1.2510e2 \\
\hline
\hline
\end{tabular}
\end{center}
\caption{Energy shifts with the spectator atom in a $P$ state with $F=1$:
numerical values of the \vdw{} shift to the energy difference
$\Delta\epsilon_+^{\left(\mathrm{\Rmnum{2}}\right)}$ between the symmetric
superpositions 
$\left(1/\sqrt{2}\right)\left(| (0,1,0)_A \, (1,1,1)_B \rangle+
| (1,1,1)_A \, (0,1,0)_B \rangle\right)$ and 
$\left(1/\sqrt{2}\right)\left(| (0,0,0)_A \, (1,1,1)_B \rangle+
| (1,1,1)_A \, (0,0,0)_B \rangle\right)$, and of
the energy difference $\Delta\epsilon_-^{\left(\mathrm{\Rmnum{2}}\right)}$
between the antisymmetric superpositions 
$\left(1/\sqrt{2}\right)\left(| (0,1,0)_A \, (1,1,1)_B \rangle-
| (1,1,1)_A \, (0,1,0)_B \rangle\right)$ and
$\left(1/\sqrt{2}\right)\left(| (0,0,0)_A \, (1,1,1)_B \rangle-
| (1,1,1)_A \, (0,0,0)_B \rangle\right)$; 
as a function of the interatomic separation $R$. All
energies are given in units of the hyperfine splitting constant $\calH$
defined by (\ref{eq:HFSplit}). \label{tab:Epsilon2}}
\end{table}
\end{center}
 
\def\arraystretch{1.25}
\begin{table} [t]
\begin{center}
\begin{center}
\begin{tabular}{c@{\hspace{0.3cm}}S[table-format=-1.4e-1]S[table-format=-1.4e-1]}
\hline
\hline
R & 
\multicolumn{1}{c}{$\Delta E_+^{\left(\mathrm{\Rmnum{1}}\right)}$} &
\multicolumn{1}{c}{$\Delta E_-^{\left(\mathrm{\Rmnum{1}}\right)}$} \\
\hline
$\infty$   & 0          &  0         \\
$750\,a_0$ & -4.7272e-2 &  1.9331e-2 \\
$500\,a_0$ & -2.9284    & -1.8165    \\
$250\,a_0$ &  2.4319e1  & -2.9082    \\
\hline
\hline
\end{tabular}
\end{center}
\caption{Energy shifts with the spectator atom in an $S$ state with $F=0$:
numerical values of the \vdw{} shift to the energy difference $\Delta
E_+^{\left(\mathrm{\Rmnum{1}}\right)}$ between the symmetric superposition
$\left(1/\sqrt{2}\right)\left(| (0,1,0)_A \, (0,0,0)_B \rangle+
| (0,0,0)_A \, (0,1,0)_B \rangle\right)$ and $| (0,0,0)_A \, (0,0,0)_B \rangle$, and of the
energy difference $\Delta E_-^{\left(\mathrm{\Rmnum{1}}\right)}$ between the
antisymmetric superposition 
$\left(1/\sqrt{2}\right)\left(| (0,1,0)_A \, (0,0,0)_B \rangle-
| (0,0,0)_A \, (0,1,0)_B \rangle\right)$ and 
$| (0,0,0)_A \, (0,0,0)_B \rangle$; 
as a function of the interatomic separation $R$. The
energies are given in units of the hyperfine splitting constant $\calH$
defined by (\ref{eq:HFSplit}). \label{tab:E1}}
\end{center}
\end{table}
 
\begin{center}
\def\arraystretch{1.25}
\begin{table} [t]
\begin{center}
\begin{tabular}{c@{\hspace{0.3cm}}S[table-format=-1.4e-1]S[table-format=-1.4e-1]}
\hline
\hline
R & 
\multicolumn{1}{c}{$\Delta E_+^{\left(\mathrm{\Rmnum{2}}\right)}$} &
\multicolumn{1}{c}{$\Delta E_-^{\left(\mathrm{\Rmnum{2}}\right)}$} \\
\hline
$\infty$   & 0        & 0        \\
$750\,a_0$ & -1.4673  & 2.1880   \\
$500\,a_0$ & -2.4855  & 1.2326e1 \\
$250\,a_0$ & 8.2368e1 & 8.2379e1 \\
\hline
\hline
\end{tabular}
\end{center}
\caption{Energy shifts with the spectator atom in a $P$ state with $F=0$:
numerical values of the \vdw{} shift to the energy difference $\Delta
E_+^{\left(\mathrm{\Rmnum{2}}\right)}$ between the symmetric superpositions
$\left(1/\sqrt{2}\right)\left(| (0,1,0)_A \, (1,0,0)_B \rangle+
| (1,0,0)_A \, (0,1,0)_B \rangle\right)$ and 
$\left(1/\sqrt{2}\right)\left(| (0,0,0)_A \, (1,0,0)_B \rangle+
| (1,0,0)_A \, (0,0,0)_B \rangle\right)$, and of the energy
difference $\Delta E_-^{\left(\mathrm{\Rmnum{2}}\right)}$ between the
antisymmetric superpositions 
$\left(1/\sqrt{2}\right)\left(| (0,1,0)_A \, (1,0,0)_B \rangle-
| (1,0,0)_A \, (0,1,0)_B \rangle\right)$ and
$\left(1/\sqrt{2}\right)\left(| (0,0,0)_A \, (1,0,0)_B \rangle-
| (1,0,0)_A \, (0,0,0)_B \rangle\right)$; 
as a function of the interatomic separation $R$. All
energies are given in units of the hyperfine splitting constant $\calH$
defined by (\ref{eq:HFSplit}). \label{tab:E2}}
\end{table}
\end{center}

%
%
\section{Conclusions}
\label{conclu}

We analyze the $(2S; 2S)$ interaction at the dipole-dipole level with respect to 
degenerate subspaces of the hyperfine-resolved unperturbed Hamiltonian.
Full account is taken of the manifolds with $n=2$ and $J=1/2$ ($2S$ and $2P_{1/2}$ states),
while the fine-structure splitting is supposed to be 
large against the \vdw{} energy shifts ($2P_{3/2}$ state not included in the treatment).

We find that the total Hamiltonian given in Eq.~\eqref{H} commutes with the
magnetic projection $F_z$ of the total angular momentum of the two atoms.
Hence, we can separate the manifolds with $n=2$ and $J=1/2$ into submanifolds
with $F_z=+2,1,0,-1,-2$.  In each of these manifolds, we can identify two
irreducible submanifolds, uncoupled to one another because of the usual
selection rules of atomic physics. In each of these submanifolds the
Hamiltonian matrix can readily be evaluated [see
Eqs.~\eqref{matFz2Irr1},~\eqref{matFz2Irr2},~\eqref{matFz1Irr1},~\eqref{matFz1Irr2},~\eqref{matFz0Irr1},~\eqref{matFz0Irr2},~\eqref{matFzM1Irr1},~\eqref{matFzM1Irr2},~\eqref{matFzM2Irr1}
and~\eqref{matFzM2Irr2}].  Several degenerate subspaces with first-order \vdw{}
shifts [in the parameter $\calV$ defined by (\ref{eq:Inter}),
and hence, of order $1/R^3$] can be
identified. The corresponding shifts are of course the relevant ones for large
interatomic separations.

However, it should be noted that 
those hyperfine transitions where both atoms are in 
$S$ states, actually undergo only second-order 
\vdw{} shifts, where the energy shifts are 
given by expressions proportional to $\calV^2/\calL$,
with $\calV$ being defined in Eq.~\eqref{eq:Inter}.
The relevant states and energy shifts are given in 
Eqs.~\eqref{stateFz1M1S1},~\eqref{stateFz1M1S2},~\eqref{eps13}
and~\eqref{eps24} (for the $F_z = 1$ manifold).
For the $F_z = 0$ manifold, we have the states 
given in Eqs.~\eqref{stateFz0M1S1},~\eqref{stateFz0M1S4},
as well as~\eqref{stateFz0M1S2} and~\eqref{stateFz0M1S5},
and the energy eigenvalues are provided in 
Eqs.~\eqref{EFz0M1S1},~\eqref{E24}, and~\eqref{E356}.
The transitions are labeled 
$| \Psi_{1}^{\left(\mathrm{\Rmnum{1}}\right)} \rangle \to
| \Psi_{4}^{\left(\mathrm{\Rmnum{1}}\right)} \rangle$
and
$| \Psi_{2}^{\left(\mathrm{\Rmnum{1}}\right)} \rangle \to
| \Psi_{5}^{\left(\mathrm{\Rmnum{1}}\right)} \rangle$
in Sec.~\ref{sec4}.
Experimentally, the states with both atoms in an $S$ level are 
most interesting, because they are the only ones 
that survive for an appreciable time in an atomic 
beam; $P$ states (and thus, states with $P$ admixtures)
decay with typical 
lifetimes on the order of $10^{-8} \, {\rm s}$
(see Ref.~\cite{BeSa1957}).

\begin{figure*}[t]
\begin{center}
\begin{minipage}{0.91\linewidth}
\begin{center}
\includegraphics[width=0.8\linewidth]{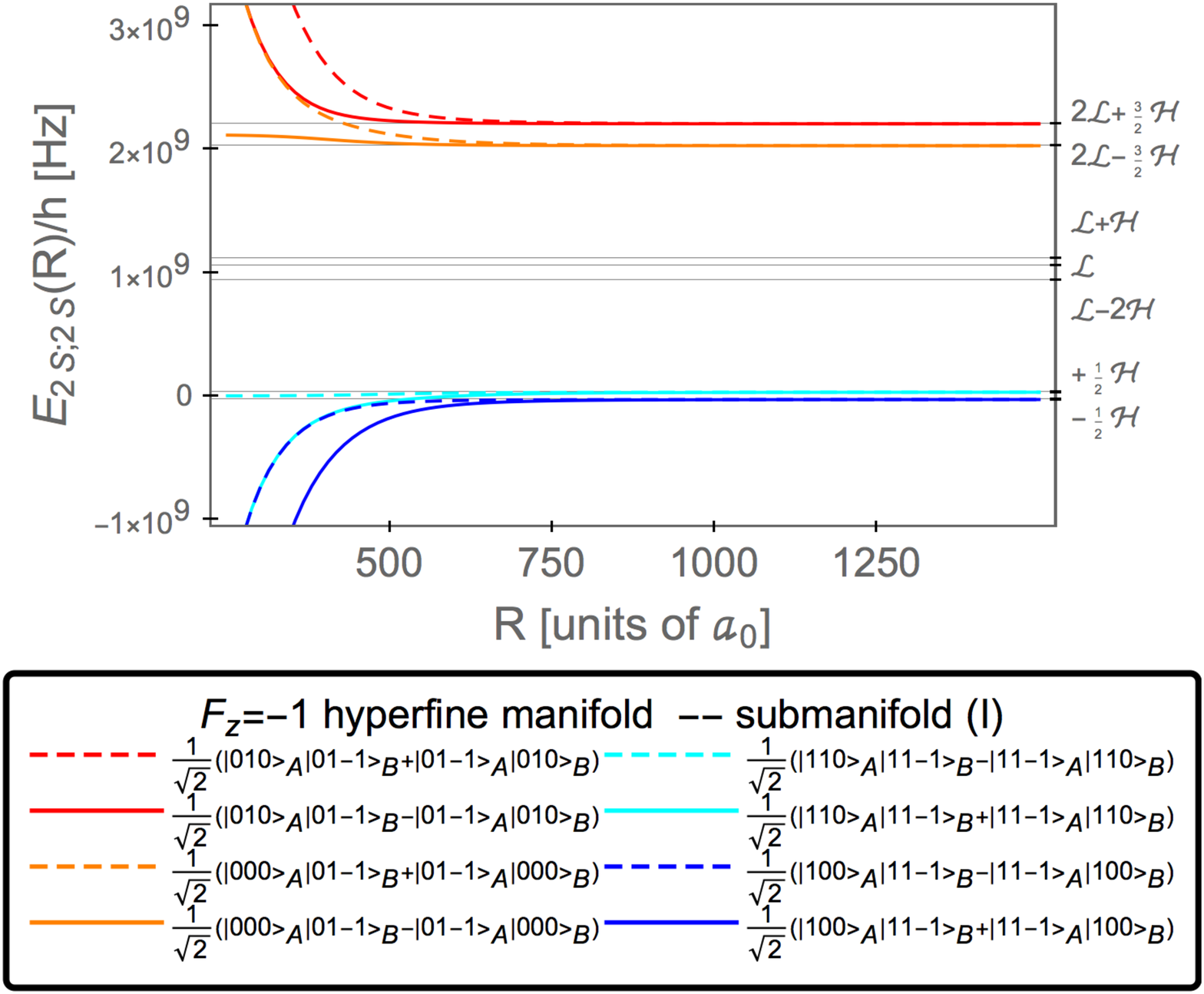}
\caption{(Color online.) Evolution of the energy levels of the submanifold
$\mathrm{\Rmnum{1}}$ within the $F_z=-1$ hyperfine manifold as a function of
interatomic separation.  The eigenstates given in the legend are only
asymptotic, for finite separation these states mix. \label{fig9}}
\end{center}
\end{minipage}
\end{center}
\end{figure*}

\begin{figure*}[t]
\begin{center}
\begin{minipage}{0.91\linewidth}
\begin{center}
\includegraphics[width=0.8\linewidth]{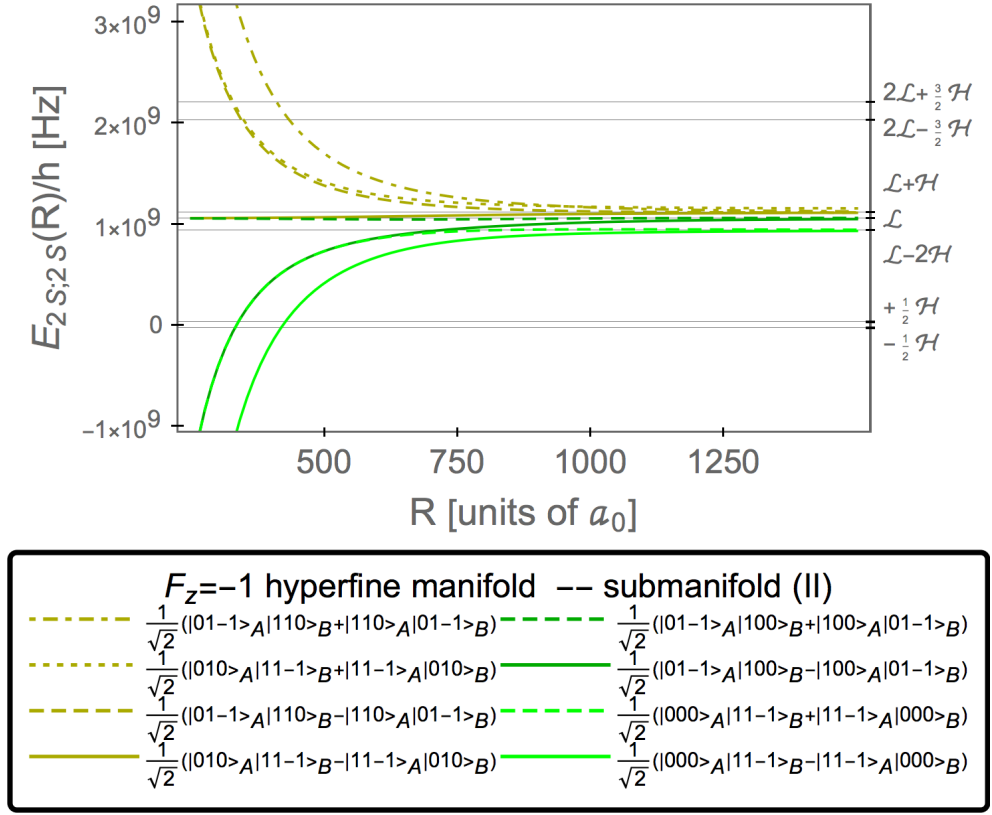}
\caption{(Color online.) Evolution of the energy levels of the submanifold
$\mathrm{\Rmnum{2}}$ within the $F_z=-1$ hyperfine manifold as a function of
interatomic separation.  The eigenstates given in the legend are only
asymptotic, for finite separation these states mix. The curve for the seventh
state in the legend (counted from the top) has been slightly offset for better
readability, in actuality it is virtually indistinguishable from that for the
sixth state.
\label{fig10}}
\end{center}
\end{minipage}
\end{center}
\end{figure*}

\begin{figure*}[t]
\begin{center}
\begin{minipage}{0.91\linewidth}
\begin{center}
\includegraphics[width=0.8\textwidth]{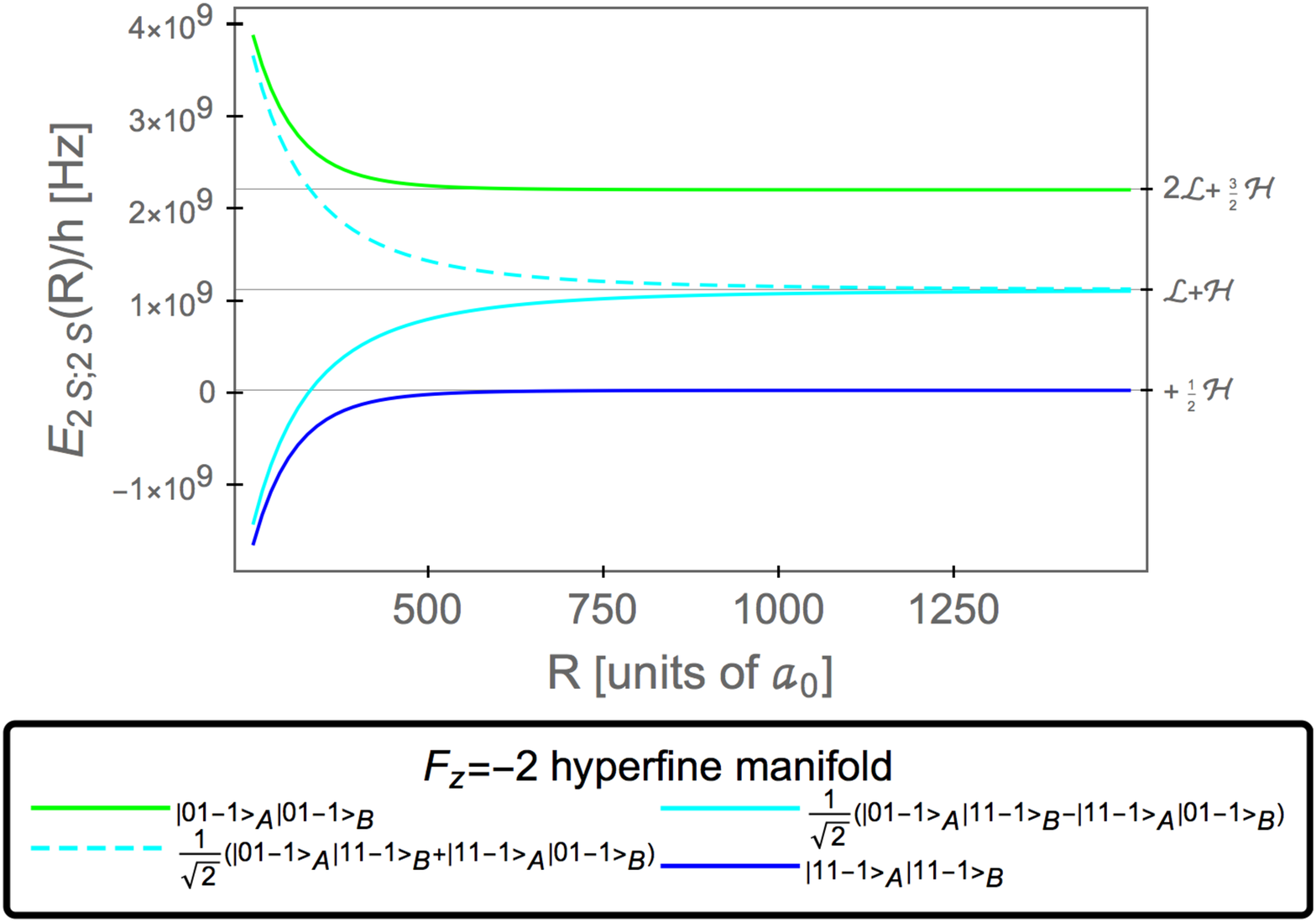}
\caption{(Color online.) Evolution of the energy 
levels within the $F_z=-2$ hyperfine manifold
as a function of interatomic separation. The eigenstates given in the legend
are only asymptotic, for finite separation these states mix.
\label{fig11}}
\end{center}
\end{minipage}
\end{center}
\end{figure*}

The dipole-dipole interaction results in level crossings (see
Figs.~\ref{fig4}---\ref{fig10}), which is a feature of the
hyperfine-resolved treatment of the problem. We are able to confirm that, in
the coarse-structure limit $\calL\rightarrow0$,
$\mathcal{F}\rightarrow0$, no such level crossings are present (as found in
Ref.~\cite{JoEtAl2002}). We note that, in the hyperfine resolved problem, there
are no level crossings for the $F_z=\pm2$ manifolds (see
Figs.~\ref{fig1} and~\ref{fig11}); for the $F_z=\pm1$ manifolds,
only crossings between levels belonging to different irreducible submanifolds
take place (in other words, the energies of states which are asymptotically of
the $2S$--$2P$ type on the one side, and of states of the $2S$--$2S$ and
$2P$--$2P$ type on the other, cross  for $R<500\,a_0$, see
Figs.~\ref{fig2},~\ref{fig3},~\ref{fig9}
and~\ref{fig10}); while, for the $F_z=0$ manifold,  both intra-submanifold (for
$R<1\,000\,a_0$) and inter-submanifold (for $R<500\,a_0$) level crossings take
place (see Figs.~\ref{fig4},~\ref{fig5},~\ref{fig6},~\ref{fig7} and~\ref{fig8}).

Of particular phenomenological interest are the $2S$ hyperfine singlet to
hyperfine triplet transitions with $| (0,0,0)_A \rangle \to | (0,1,0)_A
\rangle$ with the spectator atom $B$ in a specific state.  We find that all
transitions with the spectator atom in a $P$ state undergo first-order \vdw{}
shifts (of order $1/R^3$), while the shift is of order $1/R^6$ if the spectator
atom is in an $S$ state, that is, of second order in $\calV$. This is due
to the fact that $2S$--$2S$ states are not coupled to energetically degenerate
states (they are only coupled to $2P$--$2P$ states), while $2S$--$2P$ states
are coupled to $2P$--$2S$ states with which they are energetically degenerate.
In other words, these different behaviors are ultimately due to the selection
rules.  The spectator atom in a $P$ state, however, decays very fast to the
ground state by one-photon emission, with a lifetime of approximately
$1.60\times10^{-9}\,\mathrm{s}$ \cite{BeSa1957}, so that, depending on the
exact experimental setup, the large \vdw{} interaction energy shifts of the
$2S(F=0) \to 2S(F=1)$ hyperfine transition (with the spectator atom being in a
$2P$ state) do not play a role in the analysis of atomic beam experiments.
Otherwise, we observe that a spectator atom in a $P$ state
induces larger frequency shifts, comparing, e.g.,
the shifts in Tables~\ref{tab:Epsilon1}
and~\ref{tab:Epsilon2} for $R=750\,a_0$ and $R=500\,a_0$. 


As shown in Sec.~\ref{sec4},
the precise numerical coefficients of the 
\vdw{} shifts of the 
hyperfine singlet to hyperfine
triplet transitions $| (0,0,0)_A \rangle \to | (0,1,0)_A \rangle$
depend on the symmetry of the wave function
superposition of atoms $A$ and $B$, and cannot 
be uniquely expressed in terms of a specific 
state of the spectator atom $B$ alone;
a symmetrization term is required 
[see the term prefixed with $\pm$ in 
Eqs.~\eqref{eps13},~\eqref{eps24} and~\eqref{E24}, the same is true in the $F_z=-1$ subspace].
For spectroscopy, one essential piece of information to be derived 
from the results given in Eqs.~\eqref{eps13},~\eqref{eps24},~\eqref{E24} 
and~\eqref{E356} is that the \vdw{} interaction energy
shift for $2S(F=0) \to 2S(F=1)$ hyperfine transitions 
(with the spectator atom in a metastable $2S$ state)
is of order $\calV^2/\calL$, where 
the parameters are defined in Eq.~\eqref{parameters}
[see also the remark in the text following Eq.~\eqref{eq:AlphaBeta}].
It is straightforward to see from Eq.~(\ref{eq:Inter}) that, 
for interatomic separation $R\sim5\times10^{5}\,a_0\simeq2.6\times10^{-5}\,\mathrm{m}$, 
the van der Waals shift reaches the experimental 
accuracy of the $2S$ hyperfine frequency measurements \cite{KoEtAl2009}.

Expressed more conveniently, still in SI mksA units, 
the shift is of order
\begin{equation}
E_{2S;2S}(R) \sim \frac{\calV^2}{\calL} 
\sim E_h \left(\frac{a_0}{R} \right)^6 \, \frac{E_h}{\calL}
\end{equation}
where $E_h$ is the Hartree energy, $a_0$ is the Bohr 
radius, and $\calL \sim \alpha^3 \, E_h$ is the Lamb shift energy
[see Eq.~\eqref{defcalL}].

A quick word is in order about how the present results can be transposed to
hydrogen-like systems such as positronium and muonium. For positronium, the
hierarchy between the fine structure, Lamb shift and hyperfine structure is not
the same as that for hydrogen, so that the treatment used here; based on that
hierarchy, does not apply. For muonium, on the other hand, our analysis remains
relevant. Given that the reduced mass for the muonium system is very close to
that of the hydrogen atom, the fine structure and Lamb shift-type splittings
are almost identical to those of hydrogen. The hyperfine splitting is
$(g_s/g_N)\,(m_p/m_\mu) \sim3.2$ times larger than that of atomic hydrogen. Finally,
given the close proximity of the reduced masses, muonium has a Bohr radius very
close to that of hydrogen, so that the intensity of the dipole-dipole
interactions will be essentially identical, for equal separations, between two
hydrogen atoms and between two muonium atoms.

In this work as well as in the previous paper \cite{AdEtAl2016vdWi} of this series, we
have treated dipole-dipole interactions between atoms sitting in $S$ states
(though, in the present case, we had to treat the $2P_{1/2}$ state on the same
footing as $2S$, given their quasi-degeneracy). 
Finally, we should comment on the distance range for which our 
calculations remain applicable. We have used the 
nonretardation approximation in Eq.~\eqref{vdw}.
For the $2S$--$2S$ interaction via adjacent $2P_{1/2}$ 
states, retardation sets in when the phase of the atomic 
oscillation during a virtual (Lamb shift)
transition changes appreciably on the time scale
it takes light to travel the interatomic separation
distance $R$, i.e., when
\begin{equation}
\frac{R}{c} \sim \frac{\hbar}{\calL} \,.
\end{equation}
We have $R \sim \hbar c/\calL$ when $R$ is 
on the order of the Lamb shift wavelength of about $30 \, {\rm cm}$. 
The nonretardation approximation thus is valid over all distance
ranges of physical interest, for the ($2S$;$2S$)-system.

%
%
\section*{Acknowledgments}

The authors acknowledge insightful conversations with R.~N.~Lee.
The high-precision experiments carried
out at MPQ Garching under the guidance of Professor T.~W.~H\"{a}nsch
have been a major motivation and inspiration for the current theoretical
work.
This project was supported by the National Science Foundation
(Grant PHY--1403973).

\appendix

%
%
\section{Further Manifolds}

%
%
%
\subsection{Manifold $\maybebm{F_z = -1}$}
\label{appa1}

We can identify two irreducible subspaces within the $F_z=-1$ manifold: the
subspace $\mathrm{\Rmnum{1}}$ composed of the states
\begin{widetext}
\begin{equation}
\label{statesFzM1Irr1}
\begin{aligned} [b]
| \psi_{1}'^{\left(\mathrm{\Rmnum{1}}\right)} \rangle =& \; 
| (0,0,0)_A \, (0,1,-1)_B \rangle \,, \qquad
| \psi_{2}'^{\left(\mathrm{\Rmnum{1}}\right)} \rangle = 
| (0,1,-1)_A \, (0,0,0)_B \rangle \,, \qquad
| \psi_{3}'^{\left(\mathrm{\Rmnum{1}}\right)} \rangle = 
| (0,1,-1)_A \, (0,1,0)_B \rangle \,,
\\[0.0077ex]
| \psi_{4}'^{\left(\mathrm{\Rmnum{1}}\right)} \rangle =& \;
| (0,1,0)_A \, (0,1,-1)_B \rangle \,, \qquad
| \psi_{5}'^{\left(\mathrm{\Rmnum{1}}\right)} \rangle = \; 
| (1,0,0)_A \, (1,1,-1)_B \rangle \,, \qquad
| \psi_{6}'^{\left(\mathrm{\Rmnum{1}}\right)} \rangle = 
| (1,1,-1)_A \, (1,0,0)_B \rangle \,,
\\[0.0077ex]
| \psi_{7}'^{\left(\mathrm{\Rmnum{1}}\right)} \rangle =& \; 
| (1,1,-1)_A \, (1,1,0)_B \rangle \,, \qquad
| \psi_{8}'^{\left(\mathrm{\Rmnum{1}}\right)} \rangle = 
| (1,1,0)_A \, (1,1,-1)_B \rangle \,
\end{aligned}
\end{equation}
where the Hamiltonian matrix reads
\begin{align}
\label{matFzM1Irr1}
H_{F_z = -1}'^{\left(\mathrm{\Rmnum{1}}\right)} =& \; \left(
\begin{array}{cccccccc}
 2 \calL-\tfrac32 \calH & 0 & 0 & 0 & 0 & \calV & \calV & 2 \calV \\
 0 & 2 \calL-\tfrac32 \calH & 0 & 0 & \calV & 0 & 2 \calV & \calV \\
 0 & 0 & 2 \calL+\tfrac32 \calH & 0 & \calV & 2 \calV & 0 & \calV \\
 0 & 0 & 0 & 2 \calL+\tfrac32 \calH & 2 \calV & \calV & \calV & 0 \\
 0 & \calV & \calV & 2 \calV & -\tfrac12 \calH & 0 & 0 & 0 \\
 \calV & 0 & 2 \calV & \calV & 0 & -\tfrac12 \calH & 0 & 0 \\
 \calV & 2 \calV & 0 & \calV & 0 & 0 & \tfrac12 \calH & 0 \\
 2 \calV & \mathcal {V} & \calV & 0 & 0 & 0 & 0 & \tfrac12 \calH
\end{array}
\right) \,,
\end{align}
and the subspace $\mathrm{\Rmnum{2}}$ composed of the states
\begin{equation}
\label{statesFzM1Irr2}
\begin{aligned} [b]
| \psi_{1}'^{\left(\mathrm{\Rmnum{2}}\right)} \rangle =& \;
| (0,0,0)_A \, (1,1,-1)_B \rangle \,, \qquad
| \psi_{2}'^{\left(\mathrm{\Rmnum{2}}\right)} \rangle = 
| (0,1,-1)_A \, (1,0,0)_B \rangle \,, \qquad
| \psi_{3}'^{\left(\mathrm{\Rmnum{2}}\right)} \rangle = 
| (0,1,-1)_A \, (1,1,0)_B \rangle \,,
\\[0.0077ex]
| \psi_{4}'^{\left(\mathrm{\Rmnum{2}}\right)} \rangle =& \;
| (0,1,0)_A \, (1,1,-1)_B \rangle \,, \qquad
| \psi_{5}'^{\left(\mathrm{\Rmnum{2}}\right)} \rangle = 
| (1,0,0)_A \, (0,1,-1)_B \rangle \,, \qquad
| \psi_{6}'^{\left(\mathrm{\Rmnum{2}}\right)} \rangle = 
| (1,1,-1)_A \, (0,0,0)_B \rangle \,,
\\[0.0077ex]
| \psi_{7}'^{\left(\mathrm{\Rmnum{2}}\right)} \rangle =& \; 
| (1,1,-1)_A \, (0,1,0)_B \rangle \,, \qquad
| \psi_{8}'^{\left(\mathrm{\Rmnum{2}}\right)} \rangle = 
| (1,1,0)_A \, (0,1,-1)_B \rangle \,,
\end{aligned}
\end{equation}
where the Hamiltonian matrix reads
\begin{align}
\label{matFzM1Irr2}
H_{F_z = -1}^{\left(\mathrm{\Rmnum{2}}\right)} =& \; \left(
\begin{array}{cccccccc}
 \calL-2 \calH & 0 & 0 & 0 & 0 & \calV & \calV & 2 \calV \\
 0 & \calL & 0 & 0 & \calV & 0 & 2 \calV & \calV \\
 0 & 0 & \calL+\calH & 0 & \calV & 2 \calV & 0 & \calV \\
 0 & 0 & 0 & \calL+\calH & 2 \calV & \calV & \calV & 0 \\
 0 & \calV & \calV & 2 \calV & \calL & 0 & 0 & 0 \\
 \calV & 0 & 2 \calV & \calV & 0 & \calL-2 \calH & 0 & 0 \\
 \calV & 2 \calV & 0 & \calV & 0 & 0 & \calL+\calH & 0 \\
 2 \calV & \mathcal {V} & \calV & 0 & 0 & 0 & 0 & \calL+\calH
\end{array}
\right).
\end{align}
\end{widetext}

Surprisingly, the Hamiltonian 
matrix is a little different from the 
case with $F_z=+1$, even if one reorders the 
basis vectors accordingly.
The energy eigenvalues of course are the same.

Again within the $\mathrm{\Rmnum{1}}$ subspace there are no degenerate
subspaces with nonzero coupling, while, in the $\mathrm{\Rmnum{2}}$ subspace we
can identify degenerate states coupled to each other. The analysis carried out
in Sec.~\ref{sec32} applies here 
if we make the following substitutions:
\begin{equation} \label{eq:SubstiSubTwo}
\begin{aligned} [b]
| \psi_{1}^{\left(\mathrm{\Rmnum{2}}\right)} \rangle 
&\rightarrow| \psi_{1}^{\prime\left(\mathrm{\Rmnum{2}}\right)} \rangle\,,\;\;
| \psi_{2}^{\left(\mathrm{\Rmnum{2}}\right)} \rangle 
\rightarrow| \psi_{3}^{\prime\left(\mathrm{\Rmnum{2}}\right)} \rangle\,, \;\;
| \psi_{3}^{\left(\mathrm{\Rmnum{2}}\right)} \rangle 
\rightarrow| \psi_{4}^{\prime\left(\mathrm{\Rmnum{2}}\right)} \rangle\,,\\
| \psi_{4}^{\left(\mathrm{\Rmnum{2}}\right)} \rangle &
\rightarrow| \psi_{2}^{\prime\left(\mathrm{\Rmnum{2}}\right)} \rangle\,,\;\;
| \psi_{5}^{\left(\mathrm{\Rmnum{2}}\right)} \rangle 
\rightarrow| \psi_{5}^{\prime\left(\mathrm{\Rmnum{2}}\right)} \rangle\,,\;\;
| \psi_{6}^{\left(\mathrm{\Rmnum{2}}\right)} \rangle 
\rightarrow| \psi_{7}^{\prime\left(\mathrm{\Rmnum{2}}\right)} \rangle\,,\\
| \psi_{7}^{\left(\mathrm{\Rmnum{2}}\right)} \rangle &
\rightarrow| \psi_{8}^{\prime\left(\mathrm{\Rmnum{2}}\right)} \rangle\,,\;\;
| \psi_{8}^{\left(\mathrm{\Rmnum{2}}\right)} \rangle 
\rightarrow| \psi_{6}^{\prime\left(\mathrm{\Rmnum{2}}\right)} \rangle\,,
\end{aligned}
\end{equation}
so that we need not go over the analysis of degenerate subspaces again.
(Even when making this reordering, many off-diagonal terms have
different signs in $H_{F_z = +1}^{\left(\mathrm{\Rmnum{2}}\right)}$ and 
$H_{F_z = -1}^{\left(\mathrm{\Rmnum{2}}\right)}$. But only the couplings between
non-degenerate states have different signs, while coupling between degenerate
states remain identical. This latter point means that the analysis of
Sec.~\ref{sec32} also applies to $H_{F_z =
-1}^{\left(\mathrm{\Rmnum{2}}\right)}$.) However, for the sake of completeness
and clarity, in Figs.~\ref{fig9} and~\ref{fig10}, we plot the evolution of the
eigenvalues with respect to interatomic separation. Notice that the evolution
of the energy eigenstates is identical to the eigenstates in the $F_z=+1$
manifold.

%
%
\subsection{Manifold $\maybebm{F_z = -2}$}
\label{appa2}

We can identify two irreducible subspaces within the 
$F_z=+2$ manifold: the subspace $\mathrm{\Rmnum{1}}$ composed of the states
\begin{align}
| \phi_{1}^{\prime\left(\mathrm{\Rmnum{1}}\right)} \rangle &= 
| (0,1,-1)_A \, (0,1,-1)_B \rangle \,,\\
| \phi_{2}^{\prime\left(\mathrm{\Rmnum{1}}\right)} \rangle &= 
| (1,1,-1)_A \, (1,1,-1)_B \rangle \,,
\end{align}
where the Hamiltonian matrix reads
\begin{equation}
\label{matFzM2Irr1}
H_{F_z=-2}^{\left(\mathrm{\Rmnum{1}}\right)} = \left(
\begin{array}{cc}
 2 \calL+\tfrac32 \calH & -2 \calV \\
 -2 \calV & \tfrac12 \calH \\
\end{array}
\right) \,,
\end{equation}
and the subspace $\mathrm{\Rmnum{2}}$ is composed of the states
\begin{align}
| \phi_{1}^{\prime\left(\mathrm{\Rmnum{2}}\right)} \rangle &= 
| (0,1,-1)_A \, (1,1,-1)_B \rangle \,,\\
| \phi_{2}^{\prime\left(\mathrm{\Rmnum{2}}\right)} \rangle &= 
| (1,1,-1)_A \, (0,1,-1)_B \rangle \,,
\end{align}
where the Hamiltonian matrix reads
\begin{equation}
\label{matFzM2Irr2}
H_{F_z=-2}^{\left(\mathrm{\Rmnum{2}}\right)} = \left(
\begin{array}{cccc}
 \calL+\calH & -2 \calV \\
 -2 \calV & \calL+\calH \\
\end{array}
\right) \,.
\end{equation}%
We do not repeat the analysis of the eigensystem and refer the reader to
Sec.~\ref{sec31}. The results given there are immediately transposed to the
present case, by the simple substitution 
$| \phi_{i} \rangle\rightarrow| \phi'_{i} \rangle$. 
However, for the sake of completeness and clarity, in
Fig.~\ref{fig11}, we still plot the evolution of the eigenvalues with
respect to interatomic separation. Notice that the evolution of the energy
eigenstates is identical to the eigenstates in the $F_z=+2$ manifold.

\end{document}